\documentclass[12pt]{article}
\input jeff2.mac
\usepackage{verbatim,natbib,fullpage,color}
\usepackage{amsmath,amssymb,multirow,multicol,xcolor,url,epsfig,graphics,graphicx,bbm}
\usepackage{booktabs,epstopdf,color}
\bibpunct{(}{)}{;}{a}{,}{,}
\newcommand{\biblist}{
\bibliographystyle{chicago}
\bibliography{DVC}
}
\newcommand{\duefig}[2]{
\hbox to\hsize{\hss
    \vbox{\psfig{figure=#1,height=3.0in,width=3.6in}}
    \hss}
\vskip 0.15truein
\hbox to\hsize{\hss
    \vbox{\psfig{figure=#2,height=3.0in,width=3.6in}}
    \hss}
}

\newtheorem{lemma}{Lemma}[section]

\newcommand{\twofig}[2]{
\hbox to\hsize{\hss
    \vbox{\psfig{figure=#1,height=3.2in,width=3.1in}}\qquad
    \vbox{\psfig{figure=#2,height=3.2in,width=3.1in}}
    \hss}
}

\def\beqn{\begin{eqnarray*}}
\def\eeqn{\end{eqnarray*}}
\def\beq{\begin{eqnarray}}
\def\eeq{\end{eqnarray}}

\def\ld{\ldots}
\def\bi{\begin{itemize}}
\def\ei{\end{itemize}}
\def\ra{\rightarrow}

\def\Y{{\mbox{\boldmath $Y$}}}

\def\fh{\hat f_h}

\begin{document}

\title{\bf Partitioned Cross-Validation for  
  Divide-and-Conquer Density Estimation} 
\author{Anirban Bhattacharya and Jeffrey D.~Hart}

\date{}
\maketitle

\begin{abstract}
We present an efficient method to estimate cross-validation bandwidth
parameters for kernel density estimation in very large datasets where
ordinary cross-validation is rendered highly inefficient, both
statistically and computationally. Our approach relies on calculating
multiple cross-validation bandwidths on partitions of the data,
followed by suitable scaling and averaging to return a partitioned
cross-validation bandwidth for the entire dataset. The partitioned
cross-validation approach produces substantial computational gains
over ordinary cross-validation. We additionally show that partitioned
cross-validation can be statistically efficient compared to ordinary
cross-valida- tion. We derive analytic expressions for the
asymptotically optimal number of partitions and study its finite
sample accuracy through a detailed simulation study. We additionally
propose a permuted version of partitioned cross-validation which
attains even higher efficiency. Theoretical properties of the
estimators are studied and the methodology is applied to the Higgs
Boson dataset with 11 million observations.  

\medi
\end{abstract}

\ve
\section{Introduction} 

With dramatic advances in data acquisition and storage techniques,
modern applications routinely necessitate the analysis of massive
datasets. Accordingly, there has been a flurry of recent activity in
the analysis of {\em big data} \citep{jordan2013statistics}, with
emphasis on the {\em divide-and-conquer} strategy. Broadly speaking,
the divide-and-conquer approach splits the data into disjoint
subgroups, performs statistical analyses on all subgroups, and pools
together one or more statistics calculated from each subgroup to obtain
global estimates. Such an exercise is typically necessitated when the
statistical approach under consideration is computationally expensive
to implement on the full dataset. The data may also be too big to load
onto the memory on a single machine, or may be split across different
administrative units. Recent statistical applications of the
divide-and-conquer approach include parametric models
\citep{li2013statistical}, bag of little bootstraps
\citep{kleiner2014scalable}, kernel ridge regression
\citep{zhang2015divide}, semi-parametric heterogeneous models
\citep{zhao2014partially} and parallel MCMC for Bayesian methods
\citep{scott2016bayes,johndrow2015approximations}, among others. It
has been recently observed that divide-and-conquer procedures can
achieve minimax optimality in non/semi-parametric models
\citep{zhang2015divide,zhao2014partially}, provided the smoothing
parameters are chosen appropriately. However, theoretical
justifications for choosing smoothing parameters in a fully data
dependent fashion are yet to be developed in this context.  

In this article, we focus on kernel density estimation
\citep{Silverman86} for massive datasets. Given the linearity of
kernel density estimates, such methods are naturally amenable to
divide-and-conquer as simple averaging over partitions of the data
suffice to compute a global estimate, provided the bandwidth parameter
is specified. The choice of the kernel bandwidth is an ubiquitous
problem to which a large literature has been devoted
\citep{sheather2004density}. We specifically focus on the
cross-validation (CV) approach \citep{HallMarron} in this article. For
big datasets, the CV criterion becomes prohibitively expensive to
compute. With this motivation, we consider a partitioned
cross-validation (PCV) approach which partitions the data into
subgroups, calculates an ordinary CV bandwidth for each subgroup and
scales and averages these bandwidths to return a bandwidth for the
entire data. The idea of PCV was first put forward by \cite{Marron1},
who minimized the average of CV curves computed over different
partitions. Our approach instead separately minimizes each CV curve to
obtain a group specific CV bandwidth before averaging them. Although
the relative ordering of averaging and minimizing has a negligible
impact asymptotically, the proposed approach requires less
communication between the different partitions and is therefore more
amenable to parallelization. Moreover, since we only calculate CV
bandwidths on partitions of the data, we obtain computational gains in
orders of magnitude over ordinary CV. However, more interestingly, we
exhibit that PCV can be statistically more efficient than ordinary
CV. It is well known that the CV bandwidth converges to the (MISE)
optimal bandwidth at the notoriously slow rate of $n^{-1/10}$. We
argue that PCV can substantially improve this rate to $n^{-1/6}$. This
behavior indicates that divide-and-conquer has a fundamentally broader
statistical appeal than computational tractability alone.   

We provide a default choice for the number of subgroups under
normality which works well in practice for a wide range of
densities. While not explored empirically, we additionally describe a
model-averaging approach instead of using a fixed number of
subgroups. The finite sample efficacy of PCV over ordinary CV is
demonstrated through a number of replicated simulation studies. As a
further improvement over PCV, we present a permuted PCV (PCVP)
approach that calculates multiple PCV bandwidths on random
permutations of the data. We theoretically and empirically demonstrate
that permuted PCV can achieve substantial variance reduction compared
to PCV. In fact, it can improve the PCV rate of convergence from
  $n^{-1/6}$ to $n^{-2/11}$. The PCV and permuted PCV approaches are
applied to the Higgs 
Boson data, publicly available at the UCI machine learning repository,
with eleven million samples.

\section{Methodology}
\subsection{Partitioned cross-validation}
Suppose one observes a random sample $X_1,\ld,X_n$ from a density $f$. The
usual kernel density estimator of $f(x)$ is
$$
\hat f_h(x)=\frac{1}{nh}\sum_{i=1}^n K\left(\frac{x-X_i}{h}\right),
$$
where $K$ is an appropriate kernel function, usually a unimodal
density that is symmetric about 0 and having finite variance, and
$h$ is a positive number called the bandwidth. We are interested in
cases where $n$ is so large that $\hat f$ cannot be computed
directly. The divide and conquer solution to this problem begins by
randomly dividing the original data set up into $p$ mutually exclusive and
exhaustive subsamples of equal size.  A kernel estimate, call it $\hat 
f_{h}(\,\cdot\,|i)$, is computed from the $i$th subsample,
$i=1,\ld,p$, and an overall estimate of $f$ is the average of these
$p$ kernel estimates. Due to the linearity of the kernel estimate,
note that  
$$
\hat f_h\equiv \frac{1}{p}\sum_{i=1}^p\hat f_{h}(\,\cdot\,|i).
$$

An omnipresent problem associated with kernel estimators is that of
choosing the bandwidth $h$. A natural method of so-doing in our divide
and conquer situation is that of {\it partitioned
  cross-validation} (PCV), as proposed by \cite{Marron1}. Before going
into detail about PCV, we note at this point just two things about the
method: (i) it involves partitioning the data, as is done in divide
and conquer, and (ii) it leads to a more efficient bandwidth selector
than does ordinary, leave-one-out cross-validation. The second point is
interesting as it shows that partitioning, rather than leading to a
loss of efficiency in choosing a bandwidth, can actually lead to an
{\it increase} in efficiency.  

Before further discussion of PCV, it will be useful to review some
aspects of optimal
bandwidth choice and cross-validation. It will be assumed throughout
this paper that $f$ is square integrable and  
has two continuous derivatives everywhere.  As a loss
function we employ integrated squared error (ISE):
$$
ISE(\hat f_h,f)=\int_{-\infty}^\infty(\hat f_h(x)-f(x))^2\,dx.
$$
The optimal bandwidth $h_{n,0}$ is defined to be the minimizer of mean
integrated squared error (MISE), i.e, $MISE(\hat f_h,f)=E[ISE(\hat
  f_h,f)]$. It is well known \citep[see, e.g.,][]{Silverman86} that if
$n$ tends to $\infty$, 
\beq\label{opth}
h_{n,0} \sim D n^{-1/5}, \quad D = \left[\frac{R(K)}{R(f'')\sigma_K^4}\right]^{1/5},
\eeq
where $R(g)$ is $\int g^2(x)\,dx$ for any square integrable function
$g$ and $\sigma_K^2$ is the variance of $K$.

The leave-one-out cross-validation (CV) criterion is
$$
CV(h)=\int_{-\infty}^\infty\fh^2(x)\,dx-\frac{2}{n}\sum_{i=1}^n\fh^{i}(X_{i}),\quad
h>0, 
$$
where $\fh^{i}$ is a kernel estimate computed with the $n-1$
observations other than $X_{i}$. The CV bandwidth $\hat h$
is the minimizer of $CV(h)$. \cite{HallMarron} show that 
\beq\label{cvrate}
n^{1/10}\left(\frac{\hat h-h_{n,0}}{h_{n,0}}\right)\xrightarrow{{\cal D}}Z,
\eeq
where $Z$ is normally distributed with mean 0. This result shows that
the CV bandwidth converges to the optimum bandwidth at the notoriously
slow rate of $n^{-1/10}$. Note that in our divide
and conquer setting it is not possible to compute $\hat h$ because of
a prohibitively large value of $n$. 

Use of PCV, as proposed by \cite{Marron1}, can improve upon the slow rate
of convergence of the CV bandwidth. This result seems not to be very
well known, but has important implications for the divide-and-conquer
setting. Our version of PCV proceeds as follows:
\begin{itemize}
\item Given a random sample of size $n$, randomly partition it into
  $p$ groups of equal size.
\item Compute the usual CV bandwidth for each group, and denote these
  bandwidths by $\hat b_i$, $i=1,\ld,p$.
\item Each bandwidth $\hat b_i$ estimates an optimal bandwidth
  for sample size $n/p$, and so, as suggested by (\ref{opth}), 
bandwidths appropriate for a sample of size $n$ are $\hat 
h_i=p^{-1/5}\hat b_i$, $i=1,\ld,p$.
\item The PCV bandwidth is $\hat h_{PCV}=\sum_{i=1}^p\hat h_i/p$.
\end{itemize}
PCV results in a bandwidth that is less variable, but more biased,
than the CV bandwidth. However, it turns out that there are many
choices of $p$ such that the reduction in variance more than offsets
the increase in squared bias, resulting in a bandwidth $\hat h_{PCV}$
such that $E\left[(\hat h_{PCV}-h_{n,0})/h_{n,0}\right]^2$ converges
to 0 at a faster rate than $n^{-1/5}$, which is the corresponding rate
for the CV bandwidth $\hat h$.   

The version of PCV proposed by \cite{Marron1} averages the CV criteria
from the $p$ groups, and then chooses a bandwidth to minimize the
average criterion. We prefer our version 
since it requires no communication between groups. Results of
\cite{HallMarron} or 
\cite{scottterrell} entail that if $n$, $p$ and $n/p$ all tend to $\infty$,
then $\Var(\hat{h}_{PCV}) \sim A^* n^{-3/5} p^{-4/5}$ for a positive constant $A^*$ 
defined in \eqref{eq:Astar}, and hence 
$\Var(\hat{h}_{PCV})/h_{n,0}^2$ is asymptotic to $Ap^{-4/5}n^{-1/5}$
for $A = A^*/D^2$. This asymptotic variance is identical to
that obtained by  \cite{Marron1} for his version of PCV.

Now consider 
\beq\label{biash}
E(\hat h_{PCV})=h_{n,0}+B_{1,n,p}+B_{2,n,p},
\eeq
where
$$
B_{1,n,p}=p^{-1/5}h_{n/p,0}-h_{n,0}\quad{\rm and}\quad B_{2,n,p}=E(\hat
h_{PCV})-p^{-1/5}h_{n/p,0}.
$$
\cite{Marron1}
obtains an approximation to $B_{1,n,p}$ that is of
order $n^{-3/5}p^{2/5}$, and implicitly assumes that
$B_{2,n,p}$ is of smaller order than $B_{1,n,p}$. However, our Theorem
1 suggests that $B_{2,n,p}$ is of {\it larger}
order than $B_{1,n,p}$.  The rate of $B_{2,n,p}$ is determined by the
bias of the ordinary CV bandwidth as an estimator of
$h_{n,0}$. Because of the large variance of the CV bandwidth, this
bias is small enough to ignore in the classic theory of the CV
bandwidth (see (\ref{cvrate})), but dictates the dominant
bias term in PCV. 

\cite{scottterrell} and \cite{HallMarron} provide conditions under
which 
\beq\label{bwexpand}
\hat
h-h_{n,0}=-\frac{CV'(h_{n,0})}{CV''(h_{n,0})}+o_p\left(\frac{CV'(h_{n,0})}{CV''(h_{n,0})}\right).  
\eeq
Result (\ref{bwexpand}) suggests that $E(\hat h-h_{n,0})$ may be
approximated by applying the classic approximation of the expectation of
a ratio of random variables to $CV'(h_{n,0})/CV''(h_{n,0})$. Using the
fact that $E[CV'(h_0)]=0$, said classic approximation is 
\beq\label{expectratio}
E\left[\frac{CV'(h_{n,0})}{CV''(h_{n,0})}\right]&\approx&\Cov(CV'(h_{n,0}),CV''(h_{n,0})/M''(h_{n,0})^2\nonumber\\
&=&E\left[CV'(h_{n,0})CV''(h_{n,0})\right]/M''(h_{n,0})^2,
\eeq
where $M(h)\equiv MISE(\hat f_h,f)$. Theorem 1 provides a first order
approximation of (\ref{expectratio}).

\medi
{\bf Theorem 1.} \ {\it Suppose that $K$ is a symmetric-about-0
  density function satisfying the following: \\
\noindent (1) The first two derivatives of $K$ exist everywhere. \\
\noindent (2) As $u$ tends to infinity, both $K(u)$ and $K'(u)$ are $o(\exp(-a_1 u^{a_2}))$ for positive constants $a_1$ and $a_2$. \\
Assume also that
the first three derivatives of $f$ exist and are bounded and
continuous. Letting $\hat h$ and $h_{n,0}$ be as defined earlier in this
section, 
$$
E\left[CV'(h_{n,0})CV''(h_{n,0})\right]/M''(h_{n,0})^2=B^*n^{-2/5}+o(n^{-2/5})
$$
as $n\ra\infty$, where $B^*$ is a positive constant defined by \eqref{eq:Bstar} the Appendix.
}

\medskip
Assuming that indeed $E(\hat h)=h_{n,0}-B^*n^{-2/5}+o(n^{-2/5})$, we have 
$$
B_{2,n,p}=-B^*n^{-2/5}p^{1/5}+o(n^{-2/5}p^{1/5})
$$
as $n$, $p$ and $n/p$ tend to $\infty$. This entails that 
\beq\label{mse}
E\left[\frac{(\hat h_{PCV}-h_{n,0})}{h_{n,0}}\right]^2
\sim An^{-1/5}p^{-4/5}+Bn^{-2/5}p^{2/5},
\eeq
with $B = (B^*/D)^2$. 
The asymptotically optimal $p$ minimizes (\ref{mse}) and equals 
$(2A/B)^{5/6}n^{1/6}$.  The optimal rate of
convergence of $\hat h_{PCV}$ is $n^{-1/6}$, a substantial improvement
over the rate
of $n^{-1/10}$ for the CV bandwidth. 

The asymptotically optimal $p$ has the form $Cn^{1/6}$, where $C$
depends on $A$ and $B$. These 
constants depend on the kernel $K$ and the unknown density $f$. The
dependence on $K$ is not problematic, but dependence on
$f$ potentially is. However, our experience is that the latter
dependence is not a big problem. The range of choices for $p$ that lead
to an improvement over ordinary 
CV is so large that it is not difficult to find a value of $p$ that
works reasonably well.  We have had success using a ``normal reference''
choice for $C$. In other words, we use the value of $C$ for the case 
where $f$ is normal. This value, call it $C_N$, is
5.51. Interestingly, $C$ is invariant to the location 
and scale of $f$, so $C$ is parameter-free for any specified
location-scale family. Although $C_N$ is not generally optimal, we
have found that it is usually close enough to optimal to deliver a
substantial 
improvement over ordinary CV. Furthermore, since $C_Nn^{1/6}$ has the
correct rate, it {\it will} deliver an asymptotic improvement over CV.

It is conceivable that $n$ is so large that a kernel estimate cannot be
computed from a sample size of $C_Nn^{1/6}$.  In such a case one could
simply take $p$ so that $n/p$ is the largest sample size that still allows
computation of the kernel estimate.
\subsection{Using more partitions}
It seems unsatisfactory that the PCV bandwidth should be determined by
a particular ordering of the data. In principle one could
determine all possible partitions (for given $p$), compute a PCV
bandwidth for each partition, and then average the resulting
bandwidths. This idea was put forward in the article of
\cite{Marron1}. Of course, there are far too many partitions for this
to be feasible in practice.  Instead one may choose some
manageable number of random partitions. Choosing a precise
number of partitions is not so important since any number of
partitions greater than 1 will result in a bandwidth 
with smaller asymptotic variance than that of a PCV bandwidth based on a single
partition. As stated in Theorem 2 below, if one uses a number of
partitions equal to $N$ at every sample size $n$, then the asymptotic
variance of the average of the $N$ PCV bandwidths is $AV_1/N$,
where $AV_1$ is the asymptotic variance of $\hat h_{PCV}$. However,
there is a limit to the rate at which the variance can tend to 0, and
the limit is $AV_1/p$. This asymptotic variance is attained when $N$
tends to infinity and is of larger order than $p$. If $N$ is taken
to be $rp$ for a constant $r$, then the asymptotic variance is
$(1+r^{-1})AV_1/p$.  

\medi
{\bf Theorem 2.} \ {\it Suppose that the first four derivatives
    of $f$ exist and are bounded and continuous. Assume also that $K$
    satisfies the conditions of Theorem 1.
  Define $\bar h_N$ to be the average of PCV bandwidths computed from
  $N$ random permutations of the data $X_1,\ld,X_n$. Then if $p$ tends
  to $\infty$ with $p=o(n^{4/9})$, 
$$
\lim_{n\ra\infty}\frac{\Var(\bar h_N)}{\Var(\hat h_{PCV})}\cdot
\left[\frac{1}{N}+\frac{(N-1)}{Np}\right]^{-1}=1. 
$$
}

\medskip
We may use Theorem 2 to determine the optimal rate at which $\bar h_N$
converges to the optimal bandwidth $h_{n,0}$. If $N$ tends to $\infty$
at a faster rate than $p$, then  
\beq\label{msePCVP}
E\left[\frac{(\bar h_{N}-h_{n,0})}{h_{n,0}}\right]^2
\sim An^{-1/5}p^{-9/5}+Bn^{-2/5}p^{2/5}.
\eeq
This implies that the asymptotically optimal choice of $p$ is
$$
p=\left(\frac{9A}{2B}\right)^{5/11}\,n^{1/11},
$$
and the optimal rate of convergence for $\bar h_N$ is
$n^{-2/11}$. Recall that the optimal rates for ordinary
cross-validation and PCV are $n^{-1/10}$ and $n^{-1/6}$,
respectively. 

\subsection{Unequal group sizes}
When PCV is used to deal with a single massive data set that makes
ordinary computation infeasible, it seems natural to partition the data set
into groups of equal size. However, in some cases (an example of which is seen
in Section 5) it may be  
impossible or impractical to use equal group sizes, and hence it is of
interest to 
determine what effect this has on PCV.  Suppose that
$p$ 
different kernel estimates and $p$ different cross-validation scores
are computed, and that the results are combined to obtain a single
density estimate.  
Let the $p$ data sets have sample sizes $n_1<\cdots<n_p$ with 
$n=\sum_{i=1}^pn_i$. Applying CV to group $i$
yields a bandwidth $\hat b_i$, which is adjusted for sample size
yielding $\hat h_i=(n/n_i)^{-1/5}\hat b_i$, $i=1,\ld,p$. The
asymptotic variance of $\hat h_i$ is proportional to
$n_i^{-1/5}n^{-2/5}$ as $n_i$ and $n$ tend to $\infty$,
$i=1,\ld,p$. The weighted average of $\hat h_1,\ld,\hat h_p$ that
minimizes asymptotic variance has weights that are inversely
proportional to variances, and hence we propose using the bandwidth
\beq\label{weightedPCV}
\hat h=\frac{\sum_{i=1}^pn_i^{1/5}\hat h_i}{\sum_{i=1}^pn_i^{1/5}}.
\eeq

To determine the asymptotic mean squared error of $\hat h$ we need to
make an assumption about how the sample sizes behave for large $n$. We
will assume that the sizes are balanced in a certain sense. Let $Q$ be
the quantile function of $Y$, a positive, absolutely continuous random
variable with finite mean $\mu$. We assume that
$$
n_i=\frac{nQ(i/p)}{\sum_{j=1}^pQ(j/p)},\quad i=1,\ld,p.
$$
It is straightforward to argue that
both the variance and bias of 
$\hat h$ depend on $n_1,\ld,n_p$ only through
$\sum_{i=1}^pn_i^{1/5}$. Defining $\mu_{1/5}=E(Y^{1/5})$, we have 
$$
\sum_{i=1}^pn_i^{1/5}\sim p^{4/5}n^{1/5}\cdot\frac{\mu_{1/5}}{\mu^{1/5}}
$$ 
as $p$ and $n$ tend to $\infty$ with $p=o(n)$. 

Finally, applying previous results, we have
$$
E\left[\frac{(\hat h-h_{MISE})}{h_{MISE}}\right]^2
\sim
An^{-1/5}p^{-4/5}\left(\frac{\mu^{1/5}}{\mu_{1/5}}\right)+Bn^{-2/5}p^{2/5}\left(\frac{\mu^{1/5}}{\mu_{1/5}}\right)^2, 
$$
where $A$ and $B$ are the same constants as in (\ref{mse}).

\subsection{Model averaging as an alternative to choosing group size
  $p$}

In Section 2.1 we suggested choosing $p$ to be the value, $C_Nn^{1/6}$, that is
asymptotically optimal when the underlying density is normal.  This is
analogous to using a normal reference bandwidth, although the choice
of $p$ is arguably not so crucial since it has only a second order
effect on the bandwidth. Nonetheless, to 
protect against the possibility that $C_Nn^{1/6}$ is far from an
optimal choice for $p$, one may use a form of model averaging wherein
bandwidths arising from different choices of $p$ are averaged.

\begin{figure}[!htbp]
\begin{center}
{\includegraphics[height=3in]{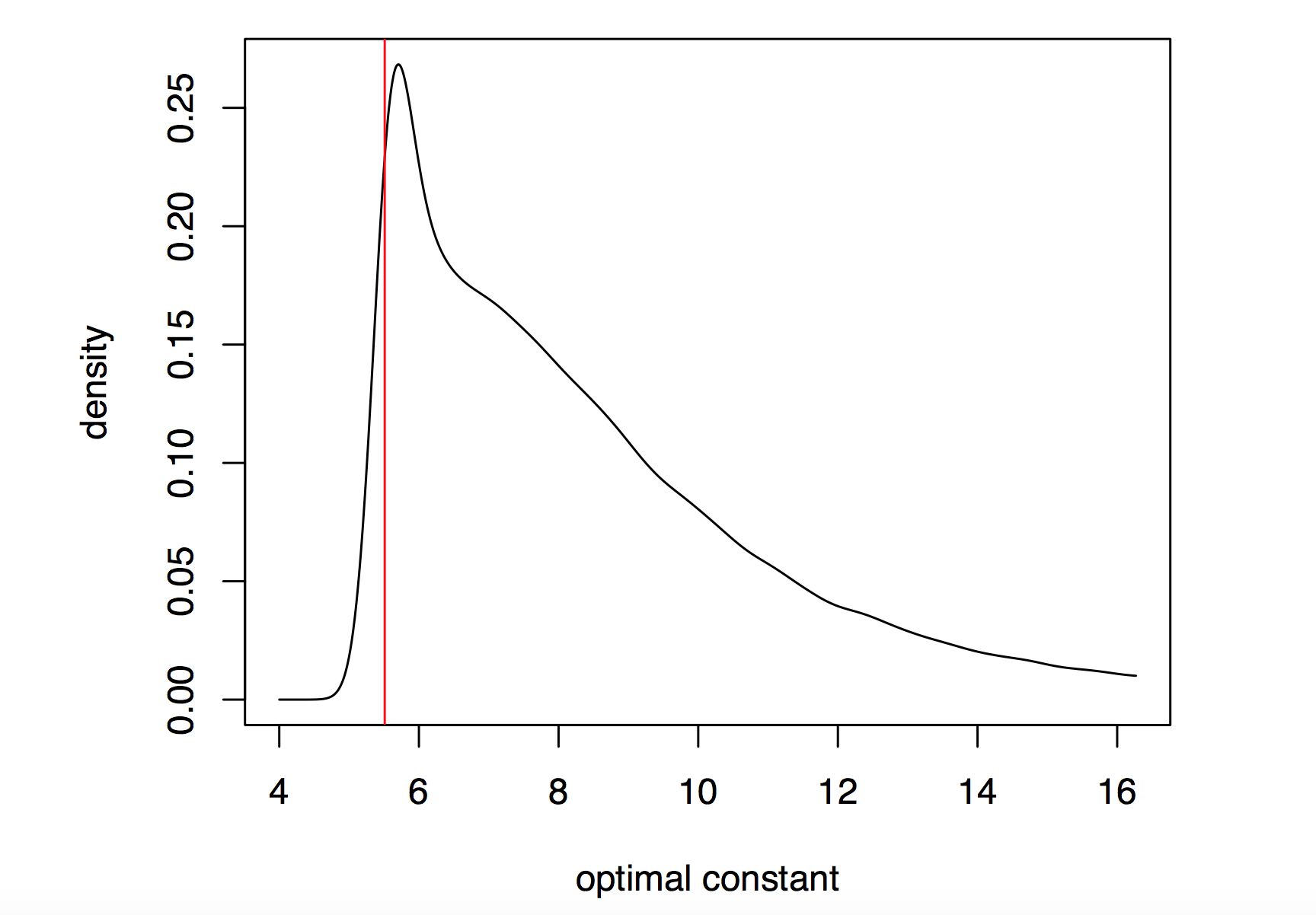}}
\caption{\it Kernel density estimate of constant in asymptotically optimal
choice of $p$. The vertical line indicates the constant of 5.51 when the
underlying density is normal.} 
\label{fig:opt_cons}
\end{center}
\end{figure}

\begin{figure}[!htbp]
\begin{center}
{\includegraphics[height=3in]{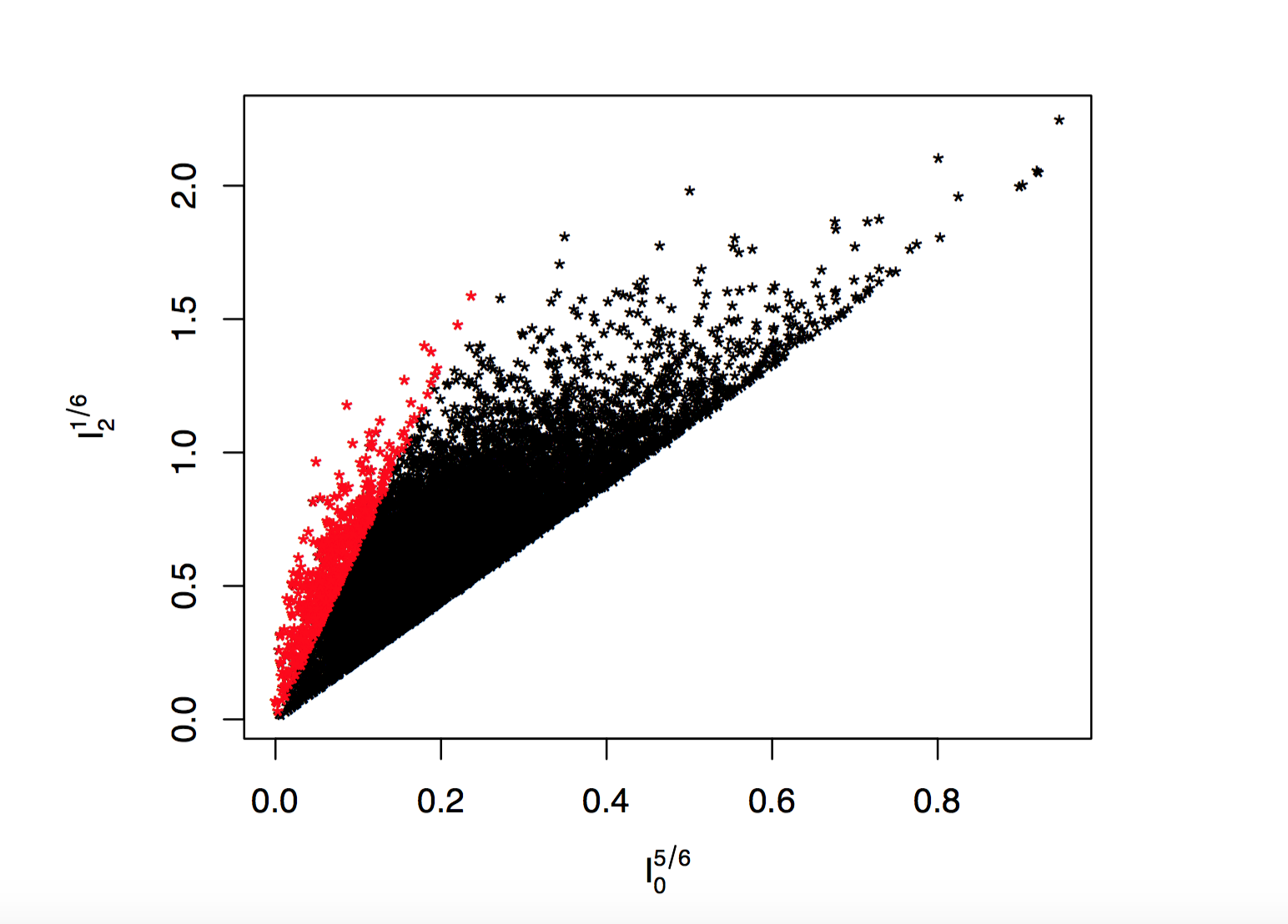}}
\caption{\it Scatterplot of $I_2^{1/6}$ versus $I_0^{5/6}$. The red
  points correspond to cases where the optimal constant is larger than
16.27.} 
\label{fig:I_0}
\end{center}
\end{figure}

The asymptotically optimal choice of $p$ has the form
$Cn^{1/6}$. To gain insight about what sort of model averaging would
be appropriate, we study how the optimal constant $C$ varies with
density $f$. As a functional of $f$, $C$ is proportional to  
\beq\label{optcon}
\frac{\left[\int (f''(x))^2\,dx\right]^{1/6}}{\left[\int
    f^2(x)\,dx\right]^{5/6}}. 
\eeq
We study the distribution of $C$ by randomly generating normal
mixtures, each of which has the form 
$$
f(x)=\sum_{i=1}^Mw_i\frac{1}{\sigma_i}\phi\left(\frac{x-\mu_i}{\sigma_i}\right). 
$$  
The densities were generated as follows:
\bi
\item A value of $M$ between $2$ and $20$ is selected from a
  distribution such that the probability of $m$ is proportional to
  $m^{-1}$, $m=2,\ld,20$. 
\item Given $m$, values $w_1,\ld,w_m$ are selected from the Dirichlet
  distribution with all parameters equal to 1/2.
\item Given $m$ and $w_1,\ld,w_m$, $1/\sigma_1^2,\ld,1/\sigma_m^2$
  are independent and identically distributed as gamma with shape and
  rate each 1/2, and conditional on $\sigma_1,\ld,\sigma_m$,
  $\mu_1,\ld,\mu_m$ are independent with $\mu_j$ distributed
  $N(0,\sigma_j^2)$, $j=1,\ld,m$.
\ei
One hundred thousand values of $C$ were obtained by generating
densities in the manner just described. The range of the 100,000
values of $C$ was $(5.4,1397.21)$. Since $C_N=5.51$, this suggests
that $C_N$ is close to being a lower bound on the optimal constant. In
Figure \ref{fig:opt_cons} we provide a kernel density estimate computed over the interval
(4,16.27), the upper endpoint of which is the 95th percentile of the
100,000 values generated. 
To see how the optimal constant correlates
with the values of $I_0=\int f^2(x)\,dx$ and $I_2=\int (f''(x))^2\,dx$, we
provide Figure \ref{fig:I_0}.  The red points in the scatterplot correspond to the
values of $C$ that were larger than 16.27. To the extent that our
generated densities represent the distribution of densities in
practice, values of $C$ larger than 16 are rare, and
extremely rare when $I_0$ is not small. 

How can the information just discussed be used to compute an average
of PCV bandwidths? Let $\hat h(p)$ be a PCV bandwidth when the
number of partitions is $p$, and let $1<p_1<p_2<\cdots<p_J<n$ be some
appropriate set of choices for $p$. Now, let $\pi$ be a prior
density for the optimal constant $C$. One possibility for $\pi$ would
be the kernel density estimate in Figure 1. Then we may define 
$$
\hat h_{\rm MA}=\frac{\sum_{j=1}^J\pi(n^{-1/6}p_j)\hat
  h(p_j)}{\sum_{j=1}^J\pi(n^{-1/6}p_j)},
$$ 
where MA stands for ``model average.'' This bandwidth is a weighted
average of PCV bandwidths, where the weight on $\hat h(p_j)$ is
the prior probability that $\hat h(p_j)$ is the (asymptotically) optimal PCV
bandwidth. Although we do not further explore this idea in the current
paper, we refer the reader to \cite{HMRV} for a discussion of the merits of
model averaging in a larger context.

\section{Simulation study}

We study various aspects of the proposed methodology through a number
of replicated simulation studies. Throughout the simulations, we use
three densities from Marron \& Wand (1992) as ground truth. Each of
these densities is a mixture of normals which facilitates comparison
as various population quantities can be calculated analytically. We
refer to these densities as MW1, MW2 and MW8 being consistent with the
numbering of Marron \& Wand (1992). The exact forms are\\ 
\noindent (i) MW1 $\equiv \mbox{ N}(0, 1)$. \\
\noindent (ii) MW2 $\equiv 0.2 \mbox{ N}(0, 1) + 0.2 \mbox{ N}(1/2,
(2/3)^2) + 0.6 \mbox{ N}(13/12, (5/9)^2)$. \\ 
\noindent (iii) MW8 $\equiv 0.75 \mbox{ N}(0, 1) + 0.25 \mbox{ N}(3/2,
(1/3)^2)$. \\ 
MW 2 and 8 respectively have a skewed unimodal and asymmetric bimodal
shape. Throughout the study we use a standard normal kernel.

We first study the finite sample accuracy of Theorem 1 in characterizing the
bias of ordinary CV bandwidths.
Letting $\tilde{h}_n = h_{n,0} - B^* n^{-2/5}$, 
Theorem \textcolor{blue}{1} suggests that $E \hat{h} = \tilde{h}_n +
o(n^{-2/5})$. We let the sample size vary from $100$ to $20000$, and
analytically calculate $h_{n,0}$ and $\tilde{h}_n$ for each sample
size. We then generate $T = 2000$ independent datasets for each sample
size and numerically minimize the CV criterion to calculate T CV
bandwidths $\hat{h}^{(1)}, \ldots, \hat{h}^{(T)}$. The Monte Carlo
estimate $\hat{h}_{MC} = (T)^{-1} \sum_{i=1}^T \hat{h}^{(i)}$ of the
CV bandwidth along with $h_{n,0}$ and $\tilde{h}_n$ are reported in
Table \ref{tab:cv}. We also calculate a $t$-statistic for testing
$E\hat{h} = \tilde{h}_n$ based on the 2000 CV samples for each sample
size. It is evident from Table \ref{tab:cv} that with increasing
sample size, the magnitude of the $t$-statistic monotonically
decreases for each of the three MW curves, taking on reasonably
moderate values for $n \ge 1000$ and being practically insignificant
for $n \ge 10000$.  

\begin{table}[!h]
\Large
\begin{flushleft} 
\scalebox{0.70}{	
\begin{tabular}{c|cccc|cccc|cccc}
\hline
\multicolumn{1}{c|}{} & \multicolumn{4}{c|}{MW1} & \multicolumn{4}{c|}{MW2} & \multicolumn{4}{c}{MW8}\\
\hline 
$n$ & Opt & Exp & MC & $t$ & Opt & Exp & MC & $t$ & Opt & Exp & MC & $t$ \\
\hline
100     & 44.55 & 41.66 &  44.06 & 8.99  & 30.53 & 28.20 &  30.69 & 9.98 & 31.79 & 30.87 & 35.15 & 14.94 \\ 
250     & 36.51 & 34.51 & 35.74 & 6.27   & 24.85 & 23.65 & 24.50  & 6.43 & 24.15 & 23.52 & 25.31 & 11.23 \\ 
500     & 31.50 & 29.99 & 30.58 & 3.86   & 21.36 & 20.45 & 20.80  & 3.33 & 20.12 & 19.64 & 20.42 & 7.22   \\ 
1000   & 27.24 & 26.09 & 26.52 & 3.50   & 18.42 & 17.72 & 17.95  & 2.70 & 16.97 & 16.61 & 16.90 & 3.62   \\ 
5000   & 19.53 & 18.92 & 19.01 & 1.14   & 13.15 & 12.79 & 12.88  & 1.79 & 11.78 & 11.58 &  11.66 & 2.06   \\ 
10000 & 16.95 & 16.49 & 16.47 & -0.42  & 11.40 & 11.13 & 11.14  & 0.25  & 10.13 & 9.99  & 10.01  & 0.70   \\ 
20000 & 14.78 & 14.38 & 14.33 & -0.92  & 9.90  & 9.69 & 9.70 & 0.29 & 8.75 & 8.64 & 8.63 & -0.62              \\ \hline
\end{tabular}
}  
\end{flushleft}
\caption{ {\it Bias of CV bandwidths for various choices of $n$ and the
  three MW curves. Opt and Exp respectively denote (100$\times$) the
  MISE optimal bandwidth $h_{n,0}$ and $\tilde{h}_n$, the
  approximation to $E \hat{h}$. MC denotes (100$\times$) the Monte
  Carlo estimate of the CV bandwidth based on $2000$ independent
  replicates and $t$ denotes a $t$-statistic based on the 2000 CV
  samples for testing $E\hat{h} = \tilde{h}_n$. }  } 
\label{tab:cv}
\end{table}

\begin{figure}
\centering
\vspace{0.05in}
\includegraphics[width=0.70\textwidth]{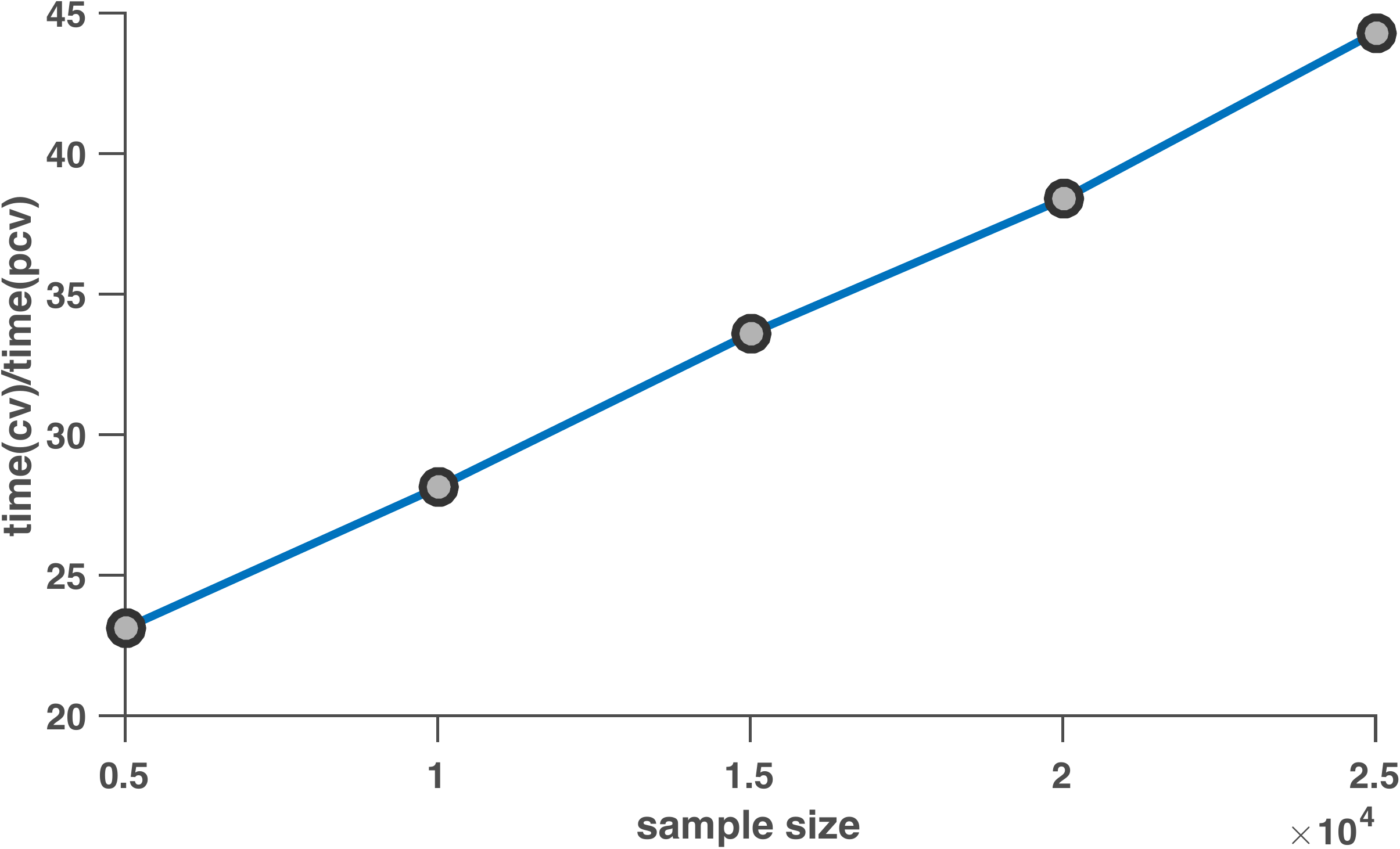}
\caption{ {\it Time comparison between ordinary CV and PCV. Ratio of computing times reported across a range of sample sizes between 5000 and 25000.} }
\label{fig:time}
\end{figure}

Our next set of simulations investigates aspects of the proposed PCV
approach. At the outset, we comment on the computational superiority
of PCV over ordinary CV. Figure \ref{fig:time} shows a plot of the
ratio of computational times between ordinary CV and PCV with
increasing sample size. The PCV was implemented with the
asymptotically optimal number of subgroups at normality for each
sample size. To avoid numerical instabilities, the computational times
for each approach were calculated by averaging over 10 datasets at
each sample size. A cross-validation function written by the authors
was used in each case to have a fair comparison. As evident from
Figure \ref{fig:time}, PCV is close to 25 times more efficient than
ordinary CV when $n = 5000$, and about 45 times more efficient when $n
= 25000$.  The computational gains would be even more pronounced for
larger values of $n$. However, for $n > 25000$, our cross-validation
function cannot be implemented due to memory and storage issues. We
should also mention that we did not take advantage of the
embarrassingly parallel nature of PCV, which would have resulted in
further time gains in the order of the number of partitions.  

\begin{figure}
\begin{flushleft}
\includegraphics[width=1\textwidth]{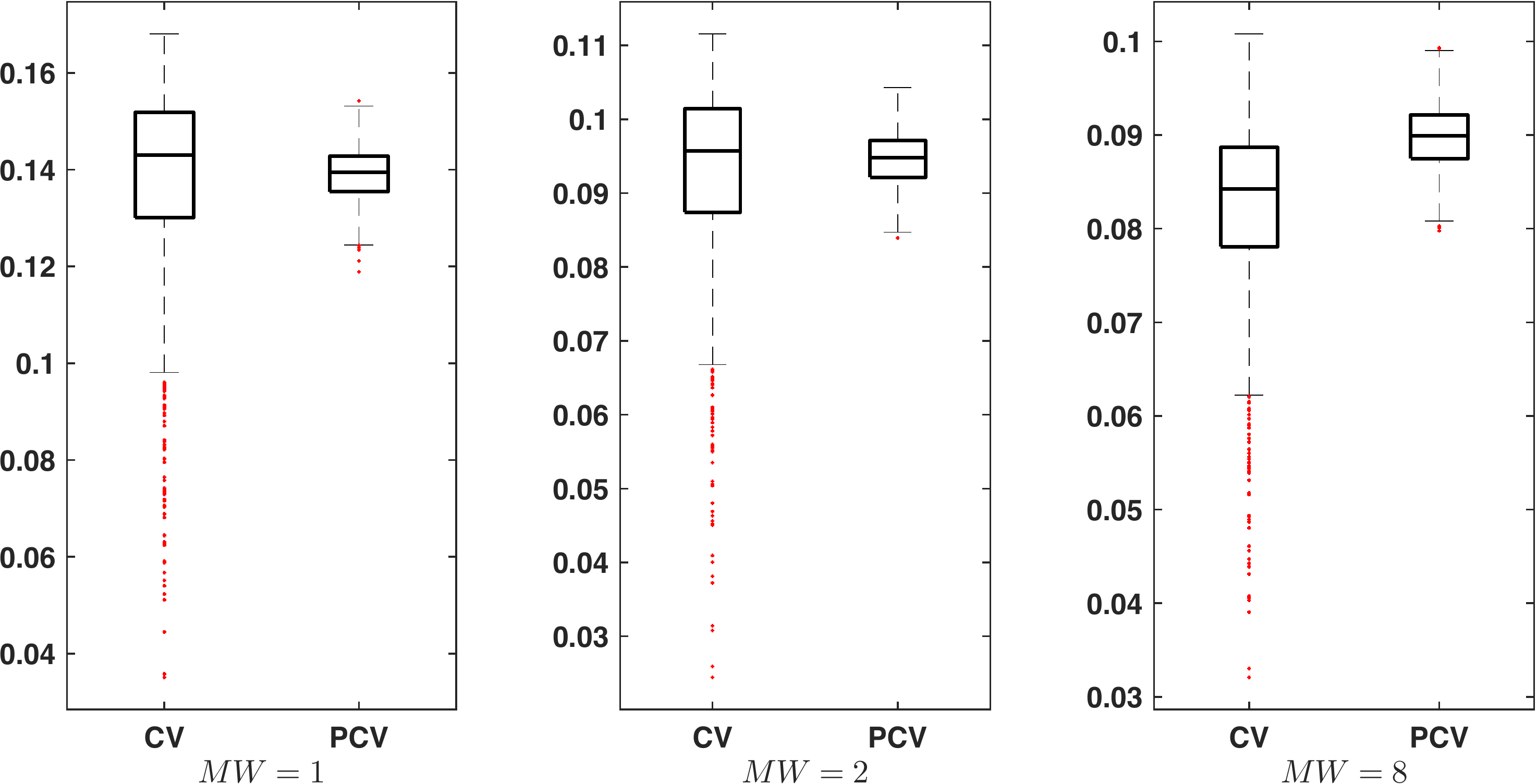}
\end{flushleft}
\caption{ {\it CV and PCV bandwidths over 1000 simulation replicates for $n
  = 25000$. The MISE optimal bandwidths are $0.140, 0.094$ and $0.084$
  for MW1, 2 and 8.} }
\label{fig:cvvspcv}
\end{figure}

We now comment on the statistical efficiency of PCV over ordinary
CV. We calculated the ordinary CV and PCV bandwidths for each of the
three MW curves over 1000 datasets at sample size $n = 25000$; box
plots of the bandwidths over the simulation replicates are provided in
Figure \ref{fig:cvvspcv}. As before, the optimal group size at
normality was used for PCV.  
The variance reduction achieved by PCV over ordinary CV is strikingly
evident in Figure \ref{fig:cvvspcv}. To quantify this variance
reduction, we compare the empirical variances with their asymptotic
counterparts. As noted in Section 2, 
\begin{align}\label{eq:var_pcv}
\Var(\hat{h}_{PCV}) \sim A^* n^{-3/5} p^{-4/5}, \quad \Var(\hat{h}_{CV}) = p^{4/5} \Var(\hat{h}_{PCV}), 
\end{align} 
where 
\begin{align}\label{eq:Astar}
A^* = \frac{8}{25} \cdot \frac{\int V^2}{(\int \phi^2)^{7/5}} \cdot \frac{\int f^2}{\{ \int (f^{''})^2 \}^{3/5} }. 
\end{align}
In the above display, $\phi$ is the standard normal density and $V$, defined in \eqref{eq:V} in the Appendix, is
determined by the kernel, with $\int V^2 = 0.0954$ for the Gaussian
kernel. When the true density $f$ is a mixture of normals as in our
case, the quantities $\int f^2$ and $\int (f^{''})^2$ can be
analytically calculated. We report the asymptotic and empirical
variances along with the variance reduction factor
$\Var(\hat{h}_{CV})/\Var(\hat{h}_{PCV})$ in Table
\ref{tab:cvvspcv}. The empirical variance reduction factor is closest
to the asymptotic approximation for the standard normal curve MW1,
with more than $15$-fold variance reduction for PCV. Even for the MW8
curve, where the asymptotic approximations seem to require a larger
sample size, PCV achieves a $9$-fold variance reduction. We also
calculated the empirical sum of squared errors $\sum_{i=1}^{1000}
\{\hat{h}^{(i)} - h_{n,0} \}^2$ for ordinary CV and PCV. PCV had
the smaller value for all three curves compared to CV, with the ratio
of CV to PCV sums of squares being $14.91, 13.09$ and
$1.98$ for the three curves respectively. The relatively smaller gain
for MW8 compared to the other two curves stems mainly from the larger bias
incurred by the PCV in this case. Specifically, for MW8 we have
$h_{n,0} = 0.0835$, with $E \hat{h}_{CV} = 0.0820$, $\Var
(\hat{h}_{CV}) = 1.01 \times 10^{-4}$, while $E \hat{h}_{PCV} =
0.0899$, $\Var (\hat{h}_{PCV}) = 1.13 \times 10^{-5}$.  
Thus, the bias somewhat offsets the variance reduction for PCV in this case. 

\begin{table}[!h]
\Large
\begin{flushleft} 
\scalebox{0.80}{	
\begin{tabular}{c|ccc|ccc|ccc}
\hline
\multicolumn{1}{c|}{} & \multicolumn{3}{c|}{MW1} & \multicolumn{3}{c|}{MW2} & \multicolumn{3}{c}{MW8}\\
\hline 
{} & CV & PCV & VRF & CV & PCV & VRF & CV & PCV & VRF \\
\hline
Asymptotic  & 29.54 & 1.95 & 15.11 & 11.89 & 0.78 & 15.11 & 54.81 & 3.62 &  15.11\\ 
Empirical     & 46.60 & 2.94 & 15.86 & 17.29 & 1.34 & 12.81 & 10.15 & 1.13 &  8.97 \\ \hline 
\end{tabular}
}  
\end{flushleft}
\caption{ {\it Variances ($ \times 10^5$) of CV and PCV bandwidths for the three MW curves with $n = 25000$. The asymptotic approximations in \eqref{eq:var_pcv} and empirical estimates from 1000 replicates are reported along with the variance reduction factor (VRF). } }
\label{tab:cvvspcv}
\end{table}

Next, we investigate the performance of PCV for larger sample sizes $n
= 5 \times 10^4, 10^5$. Due to the aforementioned difficulty with
implementing ordinary CV for large sample sizes, we use $\tilde{h}_n$
as a proxy for the CV bandwidth. Along with the asymptotically optimal
$p$ at normality, we also report the PCV bandwidth for a range of
values for $p$ in Table \ref{tab:pcv}; the reported bandwidths are
averages over 1000 simulation replicates. The last row of the table
reports the PCV bandwidth obtained by averaging over the different
choices of $p$. We picked $p = 50$ as the largest subgroup size to
ensure at least 1000 samples per subgroup. Table \ref{tab:pcv} clearly
suggests the PCV bandwidths are fairly robust with respect to choice
of the subgroup size.   

In Table
\ref{tab:pcv_var} we report the asymptotic \eqref{eq:var_pcv} and empirical estimates
for the variances of the PCV bandwidths.  For the MW1 and MW2 curves, the scenario is fairly
consistent with the $n = 25000$ case in Table \ref{tab:cvvspcv}; the
empirical variance consistently overshot the asymptotic
approximation, typically by a factor of around $1.5$. However, for the
MW8 curve, the reverse phenomenon was observed when $n = 25000$,
suggesting the necessity of larger sample sizes for the asymptotic
approximation to be accurate.  

\begin{table}[!h]
\Large
\begin{flushleft} 
\scalebox{0.85}{	
\begin{tabular}{c|cc|cc|cc}
\hline
\multicolumn{1}{c|}{} & \multicolumn{2}{c|}{MW1} & \multicolumn{2}{c|}{MW2} & \multicolumn{2}{c}{MW8}\\
\hline
{}  & $5\times 10^4$ & $10^5$ & $5 \times 10^4$ & $10^5$ & $5 \times 10^4$ & $10^5$ \\
\hline
Opt  & 12.23 & 10.63 & 8.21 & 7.14 & 7.22 & 6.27 \\ 
CV                   &  11.99 & 10.45 & 8.07 & 7.03 & 7.15 & 6.21 \\ 
PCV$_{opt}$   & 12.05 & 10.45 & 8.18 & 7.10 & 7.59 & 6.49 \\ 
PCV$_{30}$    & 12.05 & 10.45 & 8.18 & 7.08 & 7.56 & 6.45 \\ 
PCV$_{35}$    & 12.06 & 10.45 & 8.20 & 7.09 & 7.62 & 6.48 \\ 
PCV$_{40}$    & 12.07 & 10.46 & 8.21 & 7.10 & 7.66 & 6.49 \\ 
PCV$_{45}$    & 12.08 & 10.46 & 8.21 & 7.10 & 7.71 & 6.52 \\ 
PCV$_{50}$    & 12.10 & 10.49 & 8.24 & 7.12 & 7.75 & 6.54 \\ 
PCV$_{avg}$  & 12.07 & 10.46 & 8.20 & 7.10 & 7.65 & 6.50 \\ \hline
\end{tabular}
}  
\end{flushleft}
\caption{ {\it PCV bandwidths for sample sizes $n = 5 \times 10^4,
  10^5$. 1000 simulation replicates were considered. Opt and CV
  respectively denote (100$\times$) the MISE optimal bandwidth
  $h_{n,0}$ and $\tilde{h}_n$, the approximation to $E
  \hat{h}$. PCV$_{opt}$ denotes (100$\times$) PCV bandwidth using
  $p=C_Nn^{1/6}$, the optimal number of subgroups at normality, while
  PCV$_p$ used to $p$ subgroups. The value of $C_Nn^{1/6}$ is 33 at
    $n=5\times10^4$ and 38 at $n=10^5$.}  }
\label{tab:pcv}
\end{table}

\begin{table}[!h]
\Large
\begin{flushleft} 
\scalebox{0.80}{	
\begin{tabular}{c|cc|cc|cc}
\hline
\multicolumn{1}{c|}{} & \multicolumn{2}{c|}{MW1} & \multicolumn{2}{c|}{MW2} & \multicolumn{2}{c}{MW8}\\
\hline 
{} & $5\times 10^4$ & $10^5$ & $5\times 10^4$ & $10^5$ & $5\times 10^4$ & $10^5$  \\
\hline
Asymptotic & $1.17$ & $0.70$ & $0.47$ & $0.28$ & $0.22$ & $0.13$ \\ 
Empirical    & $1.74$ & $1.02$ & $0.67$ & $0.44$ & $0.62$ & $0.32$   \\ \hline 
\end{tabular}
}  
\end{flushleft}
\caption{ {\it Variances ($ \times 10^5$) of PCV bandwidths with $n = 5
  \times 10^4$ and $10^5$. The asymptotic approximations in
  \eqref{eq:var_pcv} and empirical estimates from 1000 replicates are
  reported.} }
\label{tab:pcv_var}
\end{table}

Our final set of simulations study the amount of variance reduction
achieved by permuted PCV. In the setting of Table \ref{tab:pcv}, we
also calculated the permuted PCV bandwidths with $2$ and $5$
permutations respectively. The number of subgroups was fixed at the
optimal $p$ at normality for PCV. In the following table, we report
Monte Carlo estimates of the ratio of variances between the permuted
PCV and PCV bandwidths based on 1000 datasets. The numerical results
overall agree with the conclusions of Theorem 2; one obtains a variance
reduction of approximately $1/2$ with two permutations and $1/5$ with
five. An exception is the MW8 curve for which the variance reduction
with five permutations was about $1/3$, again suggesting that the
asymptotics kick in slower for this curve. 
\begin{table}[!h]
\large
\begin{flushleft} 
\scalebox{1.00}{	
\begin{tabular}{c|cc|cc|cc}
\hline
\multicolumn{1}{c|}{} & \multicolumn{2}{c|}{MW1} & \multicolumn{2}{c|}{MW2} & \multicolumn{2}{c}{MW8}\\
\hline
method  & $5\times 10^4$ & $10^5$ & $5 \times 10^4$ & $10^5$ & $5 \times 10^4$ & $10^5$ \\
\hline
PCVP$_{2}$   & 0.51 & 0.51 & 0.50 & 0.52 & 0.55 & 0.56 \\ 
PCVP$_{5}$    & 0.24 & 0.23 & 0.23 & 0.23 & 0.33 & 0.33 \\ \hline 
\end{tabular}
}  
\end{flushleft}
\caption{ {\it Ratio of variances between permuted PCV (PCVP) and PCV
  bandwidths based on 1000 datasets. The subscript for PCVP denotes
  the number of permutations.}  }
\label{tab:pcvp_var}
\end{table}

Theorem 2 also suggests a phenomenon of diminishing returns in the
reduction of variance as the number of permutations
increases. More precisely, Theorem 2 entails that further
  reductions in variance are minimal when the number of permutations
  exceeds $p$.   
To study this, we continued with $n = 5 \times 10^4, 10^5$, and
took $p$ to be its optimal value at normality, 33 and 38 in this 
case. The number of permutations considered ranged from $1$ to $40$.
\begin{figure}[h!]
\begin{flushleft}
\includegraphics[width=0.49\textwidth]{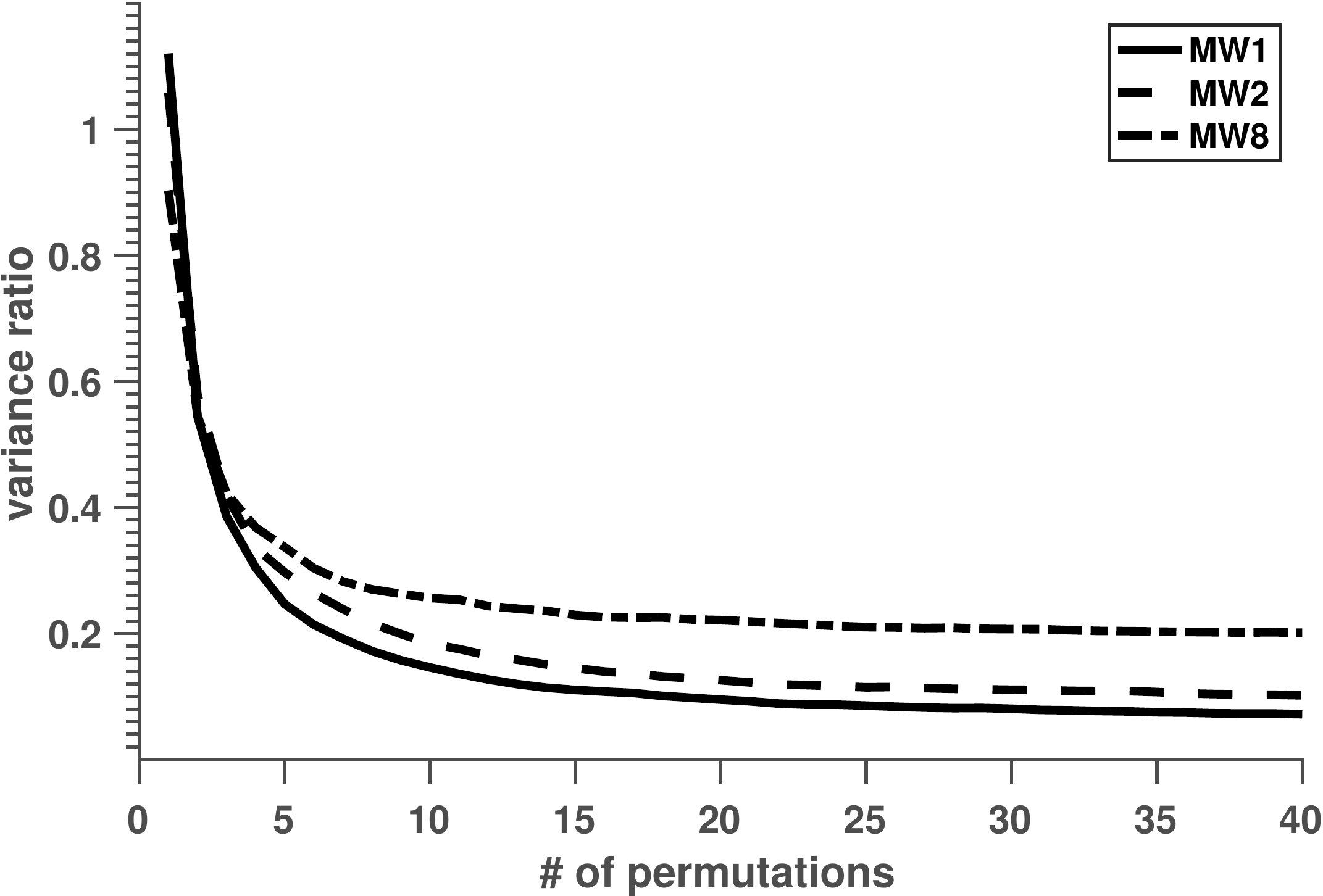}
\includegraphics[width=0.49\textwidth]{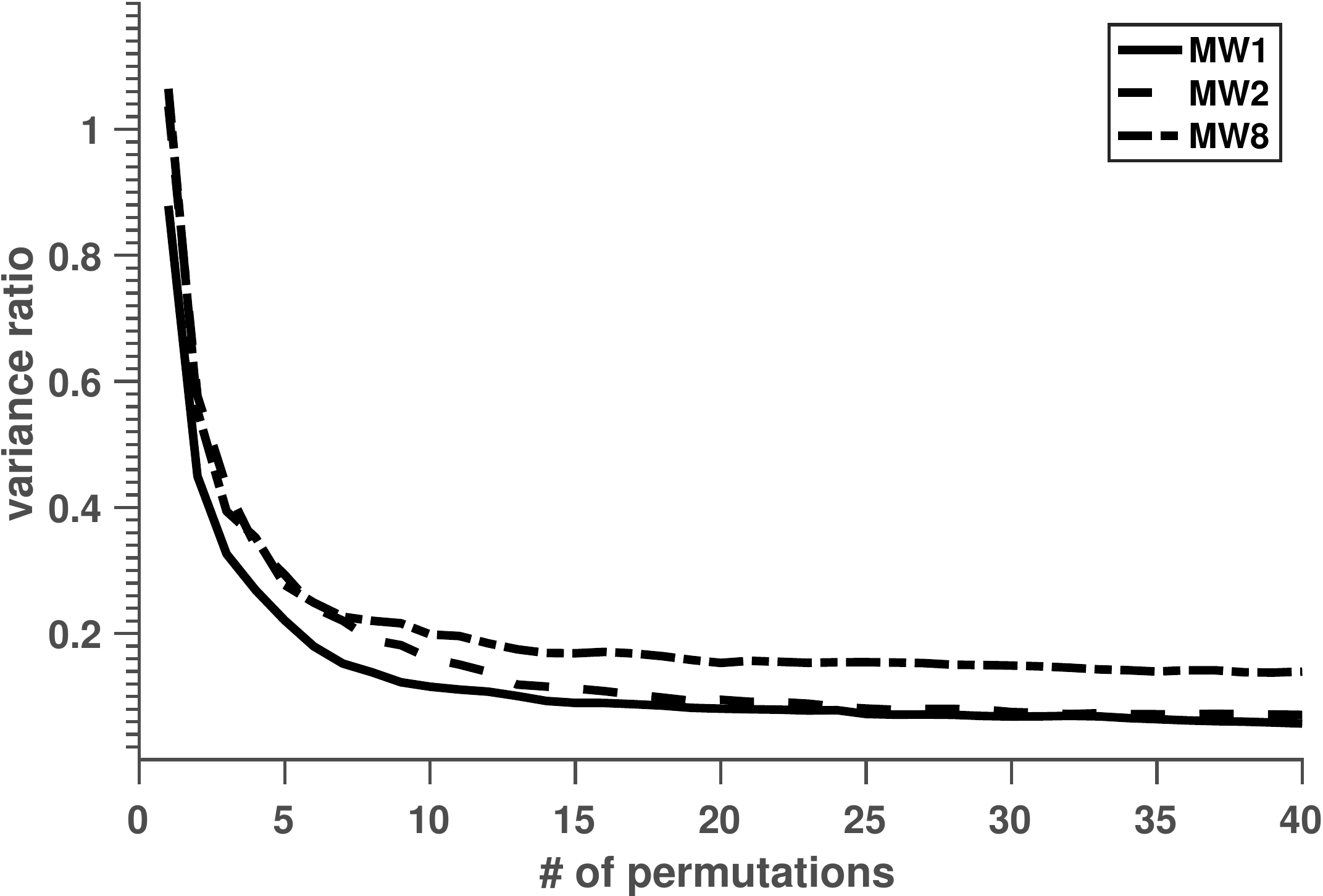}
\end{flushleft}
\caption{ {\it Ratio of variances between the permuted PCV and PCV
  bandwidths based on 1000 datasets versus the number of
  permutations. The solid/dashed/dashed-dotted curves indicate
  MW1/2/8. Sample sizes are $5 \times 10^4$ and $10^5$ in the left and
  right panels respectively.} }
\label{fig:variance}
\end{figure}
A plot of the Monte--Carlo estimates of the ratio of variances between
the permuted PCV and PCV bandwidths based on 1000 datasets against the
number of permutations is provided in Figure \ref{fig:variance}, which
clearly shows the efficacy of permuted PCV in reducing variance. In
Figure \ref{fig:variance}, we observe rapid reductions in variance
initially but then a stabilizing variance ratio beyond 15-20
permutations. For $n = 5 
\times 10^4$, the variance reduction for the MW8 curve plateaus
quickly compared to the other two curves, consistent with the
observation in Table \ref{tab:pcvp_var}. With $40$ permutations, MW8
achieves a variance reduction of about $0.20$, while the reductions 
for MW1 and MW2 are $0.05$ and $0.10$ respectively. The reductions are
more comparable for $n = 10^5$, with a reduction factor of
$0.13$ for MW8 compared to $0.05$ and $0.07$ for MW1 and MW2.

\section{Analysis of Higgs boson data}
Verifying the existence of Higgs boson is a central problem in
particle physics. Experiments are conducted in which particles collide
at very high speeds, producing exotic particles, such as the
Higgs boson. Simulated collision data have been used to study
statistical properties of various classification schemes that are
applied to collider 
data. An example of such data are the 11 million simulated
collision events studied by \cite{BSW}. These data 
may be found at the UCI Machine Learning Repository, {\tt
  archive.ics.ci.edu/ml/datasets/HIGGS}. Roughly half of the 11 
million simulated collisions are signal, meaning that they produced Higgs
bosons, and the rest are background, which produced other types of
particles. Each of the 11 million observations has 28 variables, or 
features. Here we consider just two of the 28 features, referred to as
jet 4 $\eta$ and $m_{jjj}$. (These two variables are columns 20
and 24 of the dataset at the UCI data repository.) Our goal is to
produce a total of 4 density estimates, a signal and background
estimate for each of jet 4 $\eta$ and $m_{jjj}$. 

The size of the Higgs data set presented challenges in our
analysis. It was not possible to read the entire data set into an R
session in a Linux environment. Instead, we divided the data into 110
data sets of size 100,000 each.  We then analyzed these 110 sets
separately, with not more than one data set occupying memory at the
same time. Even still there were computational issues with data sets
of size 50,000 (the typical size of a signal or background data set.)
Not enough memory could be allocated to run the cross-validation function
written by the authors, and the R function {\tt bw.ucv} always produced
a bandwidth at the upper endpoint of the interval over which the CV
curve was minimized. It was thus necessary to partition each data set
of size 50,000 into at least two subgroups in order to avoid these
computational issues.

\begin{figure}[!htbp]
\begin{center}
{\includegraphics[height=3in]{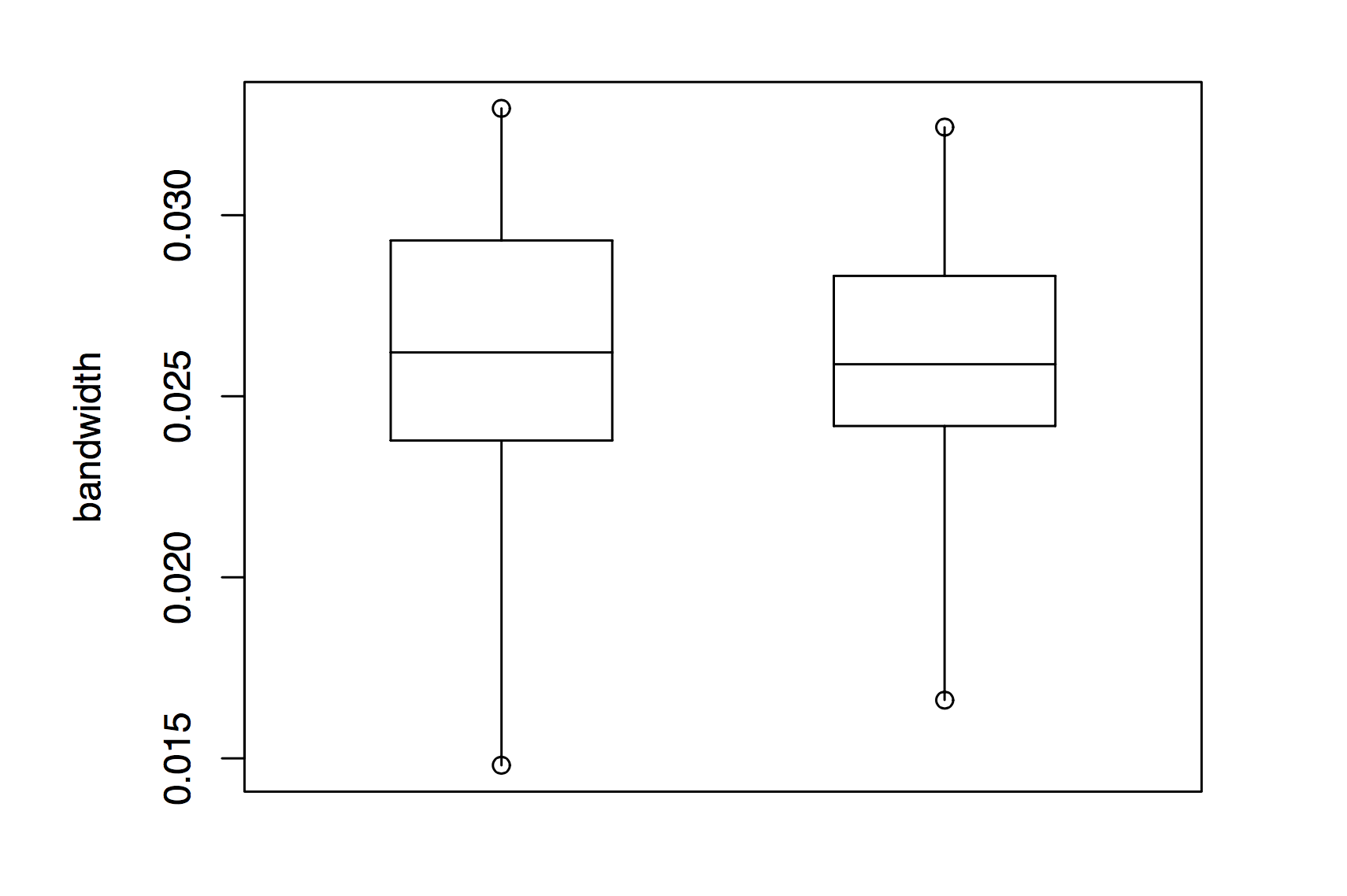}}
\caption{\it Boxplots of data-driven bandwidths for jet 4 $\eta$
  background data. The left hand plot is for 110 PCV bandwidths and the
  right hand plot for 110 permuted PCV bandwidths based on 10 random
  permutations.} 
\label{boxjet}
\end{center}
\end{figure}

Previously we suggested that the asymptotically optimal choice of $p$
at normality would be a reasonable choice of $p$ in general. For
partitioned cross-validation, with 
$n=11,000,000$ and using a Gaussian kernel, this value of $p$ is
82. In the same setting, the optimal value of $p$ for {\it permuted} PCV is
16. So, clearly we are in a situation where the sheer size of the data
set requires more partitioning than is optimal. We will apply PCV
using the smallest 
feasible number of partitions, two, for each of the 110 data sets,
yielding a $p$ of 220. While this seems far from the PCV-optimal value of
82, using 220 instead of 82 actally leads to a fairly small increase
in the mean squared error of the optimal bandwidth. When $p$ is equal
to $k$ times the optimal value of $p$, it is straightforward to
show that (\ref{mse}) takes the form  
$$
2^{-2/3}A^{1/3}B^{2/3}n^{-1/3}(k^{-4/5}+2k^{2/5}).
$$
If we use $p=220=2.68(82)$, the approximate ratio of the PCV bandwidth
MSE to the optimum MSE is thus
$[2.68^{-4/5}+2(2.68)^{2/5}]/3=1.14$. This is a fairly small increase
and suggests that the PCV bandwidth using $p=220$ will still be much
more efficient than the ordinary CV bandwidth.

\begin{figure}[!htbp]
\begin{center}
{\includegraphics[height=3in]{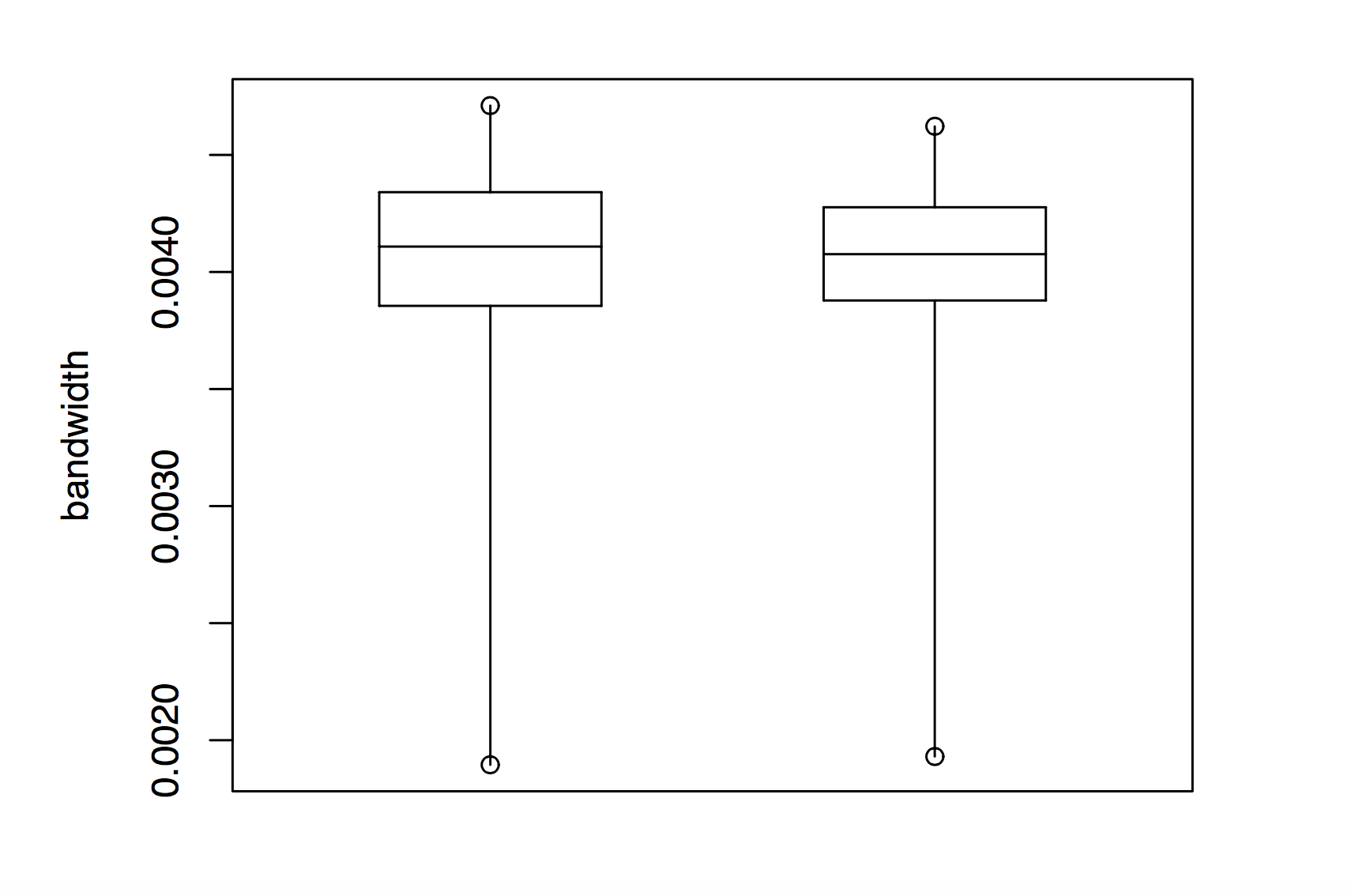}}
\caption{\it Boxplots of data-driven bandwidths for $m_{jjj}$
  background data. The left hand plot is for 110 PCV bandwidths and the
  right hand plot for 110 permuted PCV bandwidths based on 10 random
  permutations.} 
\label{boxmjjj}
\end{center}
\end{figure}

Let $n$ be the total sample size of 11 million, and let $n_{i0}$ and
$n_{i1}$ be the number of background and signal observations,
respectively, in data set $i$, $i=1,\ld,110$. For a given set of
jet 4 $\eta$ background observations, let $\hat b_{ij}$ be the
ordinary CV bandwidth for the $j$th partition of the $i$th data set,
$i=1,\ld,110$, $j=1,2$. Sample size adjusted bandwidths are $\hat
h_{ij}=n^{-1/5}(n_{i0}/2)^{1/5}\hat b_{ij}$, $i=1,\ld,110$,
$j=1,2$. Finally, the overall PCV bandwidth is defined as in
(\ref{weightedPCV}), namely
$$
\hat h=\sum_{i=1}^{110}\sum_{j=1}^2n_{i0}^{1/5}\hat
  h_{ij}\left(2\sum_{i=1}^{110}n_{i0}^{1/5}\right)^{-1}. 
$$
Applying the same procedure to each combination of feature and
background/signal led to the bandwidths shown in Table \ref{tab:boson}. 

\begin{table}
\begin{center}
\begin{tabular}{l|c|c|}
\multicolumn{1}{c}{\ }&\multicolumn{1}{c}{jet 4 $\eta$}&\multicolumn{1}{c}{$m_{jjj}$}\\
\cline{2-3}
Background&0.02588, 0.02593&0.00398, 0.00400\\
\cline{2-3}
Signal&0.02653, 0.02648&0.00667, 0.00666\\
\cline{2-3}
\end{tabular}
\caption{\it PCV bandwidths for the Higgs boson data. The left hand
  number in each cell is a PCV bandwidth, and the right hand number
  uses permuted PCV based on ten random permutations.}
\label{tab:boson}
\end{center}
\end{table}

Because of the computational issues discussed previously, we do not
permute the entire data set of 11 million observations. Instead, we
apply the permutation idea separately to the 110 smaller data sets,
and then average results.  The algorithm for a single data set is
described as follows:  
\bi
\item Compute a PCV bandwidth based on two partitions.
\item Repeat the previous step a total of ten times for ten random
  permutations of the data. 
\item Average the ten PCV bandwidths and adjust the average
  in the usual way to produce a bandwidth for a sample of size $n$. 
\ei
Having produced 110 bandwidths as described above, we then average
them to obtain the final bandwidth. The four bandwidths so determined
are given in Table \ref{tab:boson}. Obviously the bandwidths chosen by the
two methods are quite similar. An impression of the relative
variability of the methods is obtained from the boxplots in Figures
\ref{boxjet} and \ref{boxmjjj}. 
We also applied a modified version of permuted PCV to the data.
\begin{figure}[!htbp]
\begin{center}
{\includegraphics[height=3in]{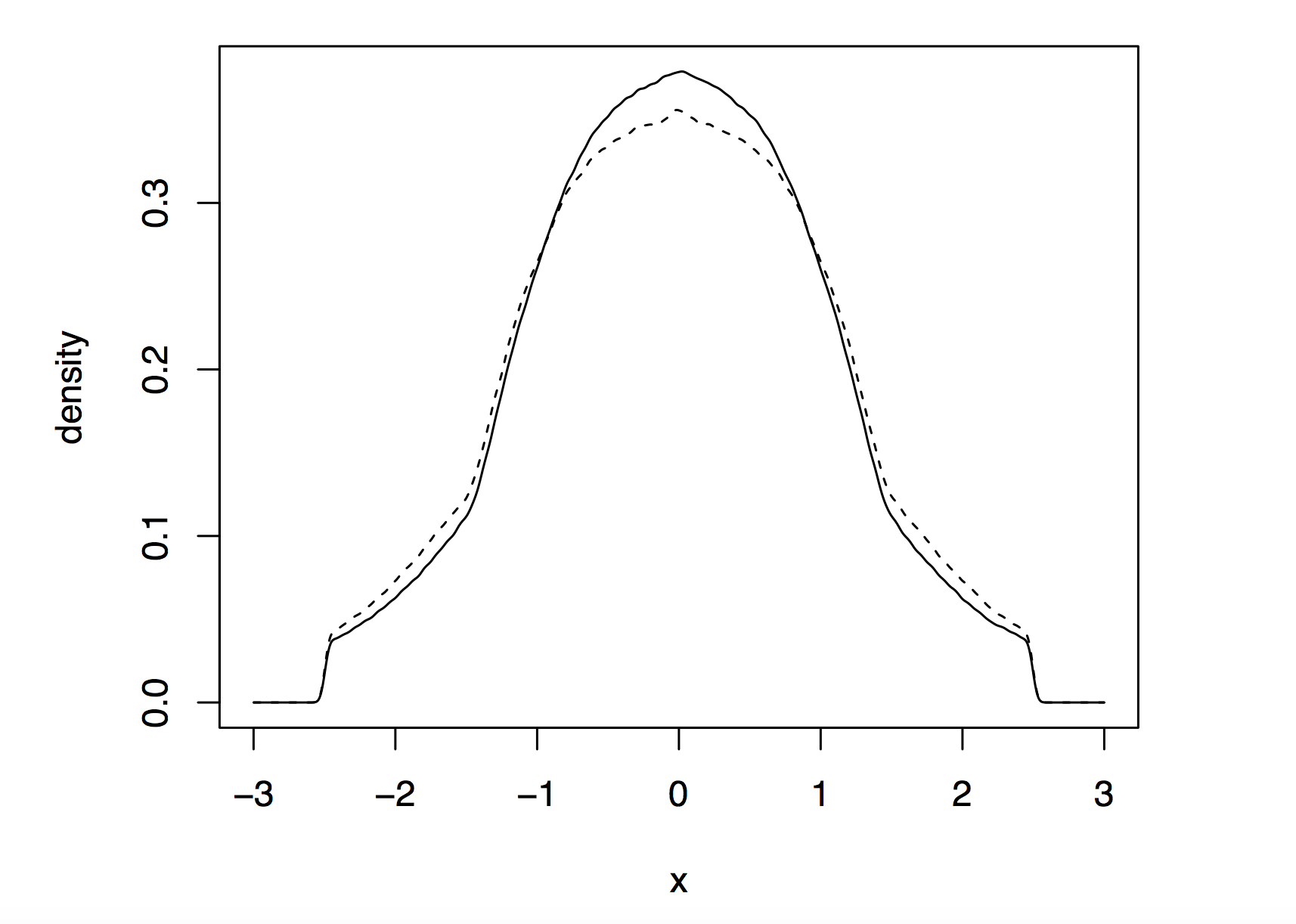}}
\caption{\it Density estimates for jet 4 $\eta$ data. The solid and
  dashed lines are for the signal and background data, respectively.} 
  \label{fig:boxjet_d}
\end{center}
\end{figure}

\begin{figure}[!htbp]
\begin{center}
{\includegraphics[height=3in]{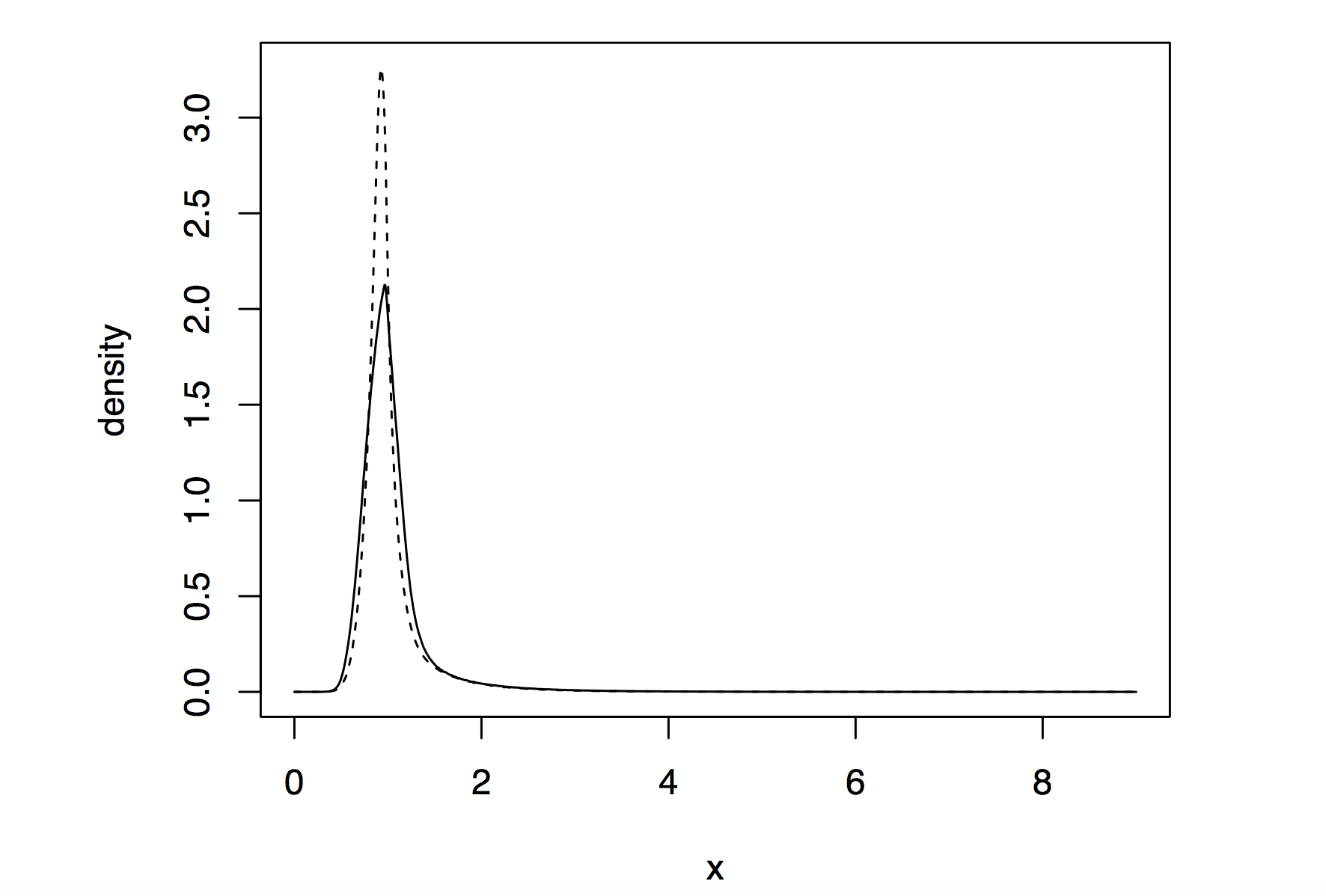}}
\caption{\it Density estimates for $m_{jjj}$ data. The solid and
  dashed lines are for the signal and background data, respectively.} 
  \label{fig:boxmjjj_d}
\end{center}
\end{figure}
Our theory indicates that were we able to consider ten random
permutations of the entire data set, then the variance of the PCV
bandwidth could be reduced by a factor of $1/10+(9/10)/220=0.104$, since we
use $p=220$ partitions. However, applying permuted PCV separately to
the 110 data sets leads to a reduction of just
$1/10+(9/10)/2=0.55$. Note that the width of the boxplots for the
permuted PCV bandwidths is about 3/4 of that for the PCV bandwidths,
which agrees with the factor of $\sqrt{0.55}=0.742$. 

Finally, we wish to produce density estimates using bandwidths from
Table \ref{tab:boson}. For a given feature, it is of interest to compare
the density estimates for signal and background. A basic principle of
comparing density estimates is to use a common bandwidth for the
estimates being compared, especially when the estimates are similar in
shape (see Bowman and Young 1996). We therefore use the
average of permuted PCV bandwidths across rows of Table \ref{tab:boson} as the
common bandwidth for signal and background. This yields bandwidths of
$0.02621$ and $0.00533$ for jet 4 $\eta$ and $m_{jjj}$, respectively. 
Separate density estimates were computed 
for the 110 data 
sets, and then a weighted average of these estimates was
computed, where the weights were proportional to sample size. In Figures
\ref{fig:boxjet_d} and \ref{fig:boxmjjj_d} we see 
that the signal and background estimates are quite similar for both
features.
This gives one an inkling of the complexity of the Higgs
classification problem since one must try to distinguish between
background and signal in a case where the marginal distributions of
these two populations are very similar for all 28 features.


\section{Concluding remarks}
Use of partitioned cross-validation to choose the bandwidth of a
kernel density estimator has been studied in the context of big
data and the divide-and-conquer scenario. It was argued that PCV
provides substantial improvements in both statistical and
computational efficiency over ordinary CV. PCV involves randomly
partitioning the data 
into $p$ subsets. Asymptotics show that the PCV bandwidth 
based on a single partitioning can converge to the optimal bandwidth at a
faster rate than does the ordinary CV bandwidth. Intuition suggests
that it is desirable to average the bandwidths resulting from multiple
random partitionings of the data. Rather remarkably, it turns out that
when the number of random  
partitionings is of a larger order than $p$, the variance of such an
average converges to 0 at a faster rate than does the variance of the
PCV bandwidth.   

Obviously partitioned cross-validation can be applied in other
contexts as well. A particularly important setting is that of
nonparametric regression, where estimation could be based on 
kernel-type estimators or splines. In either approach it is necessary
to choose smoothing parameters, and cross-validation is a commonly
used method for doing so. It seems fair to conjecture that partitioned
cross-validation will lead to improvements in statistical efficiency
in the regression context as well. 
\section{Appendix A}

{\bf Proof of Theorem 1.}  We shall use $M(h)$ to succinctly denote
$MISE(\hat{f}_h, f)$ as a function of $h$. Recall that $E \{CV(h)\} =
M(h) - \int f^2$, and $\hat{h}$ and $h_{n,0}$ are the respective
minimizers of $CV(h)$ and $M(h)$.
Define the functions $L$ and $H$ by 
\begin{align}\label{eq:der_k}
L(u) = - u K'(u), \quad H(u) = - u L'(u). 
\end{align}
Using integration by parts and the assumptions on K, $L$ and $H$ are both kernel functions
in that $\int L(u)\,du=\int H(u)\,du=1$ and $\int uL(u)\,du=\int u
H(u)\,du=0$. The kernel $H$ satisfies the further moment condition
$\int u^2 H(u)\, du=0$. Let $\tilde{f}_h$ and $f_h^*$ denote the
kernel density estimates corresponding to $L$ and $H$, i.e., 
\begin{align}\label{eq:kde_der}
\tilde{f}_h(x) = \frac{1}{nh} \sum_{i=1}^n L\bigg(\frac{x - X_i}{h}\bigg), \quad f_h^*(x) = \frac{1}{nh} \sum_{i=1}^n H\bigg(\frac{x - X_i}{h}\bigg).
\end{align}
The quantities $\tilde{f}_h^i$ and $f_h^{*i}$ are similarly defined in the usual sense.  The following identities are easily verified: 
\begin{align}\label{eq:der}
\frac{d}{dh} \hat{f}_h(x) = - \frac{1}{h} [\hat{f}_h(x) - \tilde{f}_h(x)], \quad \frac{d}{dh} \tilde{f}_h(x) = - \frac{1}{h} [\tilde{f}_h(x) - f_h^*(x)].
\end{align}

Invoking \eqref{eq:der}, 
\begin{align}
CV'(h) 
& = 2 \int \hat{f}_h(x) \frac{d}{dh} \hat{f}_h(x) dx - \frac{2}{n} \sum_{i=1}^n \frac{d}{dh} \hat{f}_h^i(x) \notag \\
& = -\frac{2}{h} \int \hat{f}_h(x) [ \hat{f}_h(x) - \tilde{f}_h(x) ] dx + \frac{2}{nh} \sum_{i=1}^n [ \hat{f}_h^i(X_i) - \tilde{f}_h^i(X_i) ] \notag \\
& = -\frac{2}{h} \int \hat{f}_h^2(x) dx  + \frac{2}{h} \int \hat{f}_h(x) \tilde{f}_h(x) dx + \frac{2}{nh} \sum_{i=1}^n \hat{f}_h^i(X_i) -  \frac{2}{nh} \sum_{i=1}^n \tilde{f}_h^i(X_i). \label{eq:cv_d1}
\end{align}
Differentiating \eqref{eq:cv_d1} and invoking \eqref{eq:der} on multiple occasions, 
\begin{align}
CV''(h) & = \frac{6}{h^2} \int \hat{f}_h^2(x)dx + \frac{2}{h^2} \int
\tilde{f}_h^2(x) dx - \frac{10}{h^2} \int \hat{f}_h(x) \tilde{f}_h(x)
dx \notag\\ 
& + \frac{2}{h^2} \int \hat{f}_h(x) f_h^*(x) dx 
 - \frac{4}{nh^2} \sum_{i=1}^n \hat{f}_h^i(X_i) + \frac{6}{nh^2}
 \sum_{i=1}^n \tilde{f}_h^i(X_i) \notag\\
& - \frac{2}{nh^2} \sum_{i=1}^n f_h^{*i}(X_i). \label{eq:cv_dd1}
\end{align}
We now introduce some further notation to express $CV'(h)$ in \eqref{eq:cv_d1} and $CV''(h)$ in \eqref{eq:cv_dd1} in a compact fashion. Define the convolutions
\begin{eqnarray}
A(x) &= \int K(x-u) K(u) du, \quad B(x) &= \int K(x-u) L(u) du  \label{eq:conv1} \\
C(x) &= \int L(x-u) L(u) du, \quad D(x) &= \int K(x-u) H(u) du. \label{eq:conv2}
\end{eqnarray}
We now record a result which expresses the individual terms appearing
in \eqref{eq:cv_d1} and \eqref{eq:cv_dd1} as U-statistics; the proof
is deferred to Appendix B.  
\begin{lemma}\label{lem:U_stat}
We have 
\begin{gather*}
\int \hat{f}_h^2(x) dx = \frac{1}{n^2h} \sum_{i=1}^n \sum_{j=1}^n A \bigg( \frac{X_i - X_j}{h} \bigg), 
\int \tilde{f}_h^2(x) dx = \frac{1}{n^2h} \sum_{i=1}^n \sum_{j=1}^n C \bigg( \frac{X_i - X_j}{h} \bigg), \\
\int \hat{f}_h(x) \tilde{f}_h(x) dx = \frac{1}{n^2h} \sum_{i=1}^n
\sum_{j=1}^n B \bigg( \frac{X_i - X_j}{h} \bigg)\  {\rm and}\\
\int \hat{f}_h(x) f_h^*(x) dx = \frac{1}{n^2h} \sum_{i=1}^n \sum_{j=1}^n D \bigg( \frac{X_i - X_j}{h} \bigg).
\end{gather*}
\end{lemma}
Using Lemma \ref{lem:U_stat} in \eqref{eq:cv_d1} and recalling the definition of $\hat{f}_h^i, \tilde{f}_h^i$, we have 
\begin{align*}
CV'(h) =  & - \frac{2}{n^2 h^2} \sum_{i=1}^n \sum_{j=1}^n \bigg[A \bigg( \frac{X_i - X_j}{h} \bigg) - B \bigg( \frac{X_i - X_j}{h} \bigg) \bigg] \\
& + \frac{2}{n(n-1)h^2} \mathop{\sum \sum}_{i \ne j} \bigg[K \bigg( \frac{X_i - X_j}{h} \bigg) - L \bigg( \frac{X_i - X_j}{h} \bigg) \bigg].
\end{align*}
Defining
\begin{align}\label{eq:V}
V(u) = A(u) - B(u) - K(u) + L(u),
\end{align}
it can be verified that 
$$
\int V(u) du = 0, \, \int u V(u) du = 0, \, \int u^2 V(u) du = 0.
$$
Noting that $-2/(n^2 h^2) = -2/\{n(n-1)h^2\} + 2/\{n^2(n-1)h^2\}$, we can write
\begin{align*}
CV'(h) = & -\frac{2}{h} \frac{1}{n(n-1)} \sum_{i=1}^n \sum_{j=1}^n \frac{1}{h} V \bigg( \frac{X_i - X_j}{h} \bigg) \\
& + \frac{2}{h} \frac{1}{n^2(n-1)} \mathop{\sum \sum}_{i \ne j} \frac{1}{h} \bigg[A \bigg( \frac{X_i - X_j}{h} \bigg) - B \bigg( \frac{X_i - X_j}{h} \bigg) \bigg].
\end{align*}
Therefore, $CV'(h) - E \{ CV'(h) \}$ may be expressed as 
\begin{align}
 &  -\frac{2}{h} \frac{1}{n(n-1)} \mathop{\sum \sum}_{i \ne j} \bigg[\frac{1}{h} V \bigg( \frac{X_i - X_j}{h} \bigg) - E_h  \bigg] \notag \\
& +  \frac{2}{h} \frac{1}{n^2(n-1)} \mathop{\sum \sum}_{i \ne j} \left\{ \frac{1}{h} \bigg[A \bigg( \frac{X_i - X_j}{h} \bigg) - B \bigg( \frac{X_i - X_j}{h} \bigg) \bigg] - E_{h1} \right\}, \label{eq:cv_d2}
\end{align}
where
\begin{align}
E_h = E \bigg[ \frac{1}{h} V\bigg( \frac{X_1 - X_2}{h} \bigg) \bigg], \quad E_{h1} = E \left\{ \frac{1}{h} \bigg[A \bigg( \frac{X_1 - X_2}{h} \bigg) - B \bigg( \frac{X_1 - X_2}{h} \bigg) \bigg] \right\}.
\end{align}
We used the fact that the terms corresponding to $i = j$ in the first sum are canceled since they are constants. 
In the expression for $CV'(h) - E \{ CV'(h) \}$ in \eqref{eq:cv_d2}, the second term in the right hand side has exactly the same form as the first with the exception that it has an extra $n^{-1}$, so that it is negligible compared to the first. We therefore conclude, 
\begin{align}
CV'(h) - E \{CV'(h)\} = -\frac{2}{h} \frac{1}{n(n-1)} \mathop{\sum \sum}_{i \ne j} \bigg[\frac{1}{h} V \bigg( \frac{X_i - X_j}{h} \bigg) - E_h  \bigg] + R_n', \label{eq:cv_d3}
\end{align}
where $R_n'$ is negligible. 

Using the same argument, we can express $CV''(h)$ in \eqref{eq:cv_dd1} in a more concise fashion as 
\begin{align}
CV''(h) - E \{CV''(h)\} = \frac{2}{h^2} \frac{1}{n(n-1)} \mathop{\sum \sum}_{i \ne j} \bigg[\frac{1}{h} W \bigg( \frac{X_i - X_j}{h} \bigg) - E_{h2}  \bigg] + R_n'', \label{eq:cv_dd2}
\end{align}
where 
\begin{align}\label{eq:W}
W(u) = 3 A(u) + C(u) - 5 B(u) + D(u) - 2 K(u) + 3 L(u) - H(u),
\end{align}
with 
$$
\int W(u) du = 0, \, \int u W(u) du = 0, \, \int u^2 W(u) du = 0, \, \int u^3 W(u) du = 0,
$$
and $R_n''$ is negligible as before. 

Ignoring terms that are negligible, we therefore have from
\eqref{eq:cv_dd2} and \eqref{eq:cv_d3} that  
\begin{align*}
& -\frac{h^3n^2(n-1)^2}{4}\Cov\{CV'(h), CV''(h)\}  \\
& = E \left\{\mathop{\sum \sum}_{i \ne j} \mathop{\sum \sum}_{k \ne l}     \bigg[\frac{1}{h} V \bigg( \frac{X_i - X_j}{h} \bigg) - E_h  \bigg]  \times \bigg[\frac{1}{h} W \bigg( \frac{X_k - X_l}{h} \bigg) - E_{h2}  \bigg] \right\} \\
& = 2n(n-1)T_1 + 4n(n-1)(n-2)T_2 + n(n-1)(n-2)(n-3)T_3,
\end{align*}
where 
\begin{align*}
T_1 &=  E \left\{ \bigg[\frac{1}{h} V \bigg( \frac{X_1 - X_2}{h} \bigg) - E_h  \bigg]  \times \bigg[\frac{1}{h} W \bigg( \frac{X_1 - X_2}{h} \bigg) - E_{h2}  \bigg]   \right\}, \\
T_2 &=  E \left\{ \bigg[\frac{1}{h} V \bigg( \frac{X_1 - X_2}{h} \bigg) - E_h  \bigg]  \times \bigg[\frac{1}{h} W \bigg( \frac{X_1 - X_3}{h} \bigg) - E_{h2}  \bigg]   \right\}, \\
T_3 &=  E \left\{ \bigg[\frac{1}{h} V \bigg( \frac{X_1 - X_2}{h} \bigg) - E_h  \bigg]  \times \bigg[\frac{1}{h} W \bigg( \frac{X_3 - X_4}{h} \bigg) - E_{h2}  \bigg]   \right\}.
\end{align*}
Note that $T_3=0$, and
\begin{align*}
T_1 
&= \int \int \frac{1}{h^2} V\bigg(\frac{x-y}{h}\bigg) W\bigg(\frac{x-y}{h}\bigg)f(x) f(y) dx dy -E_hE_{h2} \\
& = \frac{1}{h} \int V(u) W(u) du \int f^2(x) dx + o(1).
\end{align*}
Using a Taylor series expansion of $f$, the fact that $f$ has three
bounded and continuous derivatives, and $\int u^jV(u)\,du=0$,
$j=0,1,2$, it is easy to check that  
$T_2=O(h^5)$. Taking $h=h_{n,0}$ and combining results we have thus
obtained the key approximation
\begin{align}\label{eq:bias_comp}
E CV'(h_{n,0}) CV''(h_{n,0}) \sim - \frac{8}{h_{n,0}^4} \frac{1}{n^2} \int V(u) W(u) du \int f^2(x) dx.
\end{align}
Expression \eqref{eq:bias_comp} needs to be divided by
$M''(h_{n,0})^2$ in order to determine $B^*$. It suffices to replace
$M(h)$ by its well-known asymptotic expression, which we shall
continue to denote by $M$, so that  
\begin{align}\label{eq:AMISE}
M(h) = \frac{\int K^2(u) du}{nh} + \frac{\sigma_K^4 h^4 \int (f''(x))^2 dx}{4} = \frac{C_1}{nh} + C_2 h^4. 
\end{align}
Solving $M'(h) = 0$, we get $h_{n,0} \sim \{C_1/(4 C_2)\}^{1/5}
n^{-1/5}$. Differentiating $M'(h)$ and using this identity, we obtain 
\begin{align}\label{eq:AMISE_dd}
M''(h_{n,0}) \sim 5 \ 4^{3/5} C_1^{2/5} C_2^{3/5} n^{-2/5}.
\end{align}
Using \eqref{eq:AMISE_dd} and simplifying, 
\begin{eqnarray}\label{eq:Bstar}
\frac{E[CV'(h_{n,0})CV''(h_{n,0})]}{M''(h_{n,0})^2} &=& - \frac{8}{25}
\ n^{-2/5} \ \frac{ \int V(u) W(u) du \int f^2(x) dx}{\left[\int
    K^2(u)\,du\right]^{8/5}\left[\sigma_K^4\int
    (f''(x))^2\,dx\right]^{2/5} }\nonumber\\
&&+o(n^{-2/5}),
\end{eqnarray}
which defines $B^*$ and concludes the proof of Theorem 1.

\medskip

\noindent {\bf Proof of Theorem 2.}  Let $\hat h_1,\ld,\hat h_N$ be PCV
bandwidths corresponding to $N$ random permutations of the data. The
PCVP bandwidth is $\bar h=\sum_{i=1}^N\hat h_i$. We have
$$
\hat h_i=\frac{1}{p}\sum_{j=1}^p\hat h_{ij},
$$
where $\hat h_{i1},\ld,\hat h_{ip}$ are the bandwidths computed on the
$p$ groups of the $i$th partitioning, $i=1,\ld,N$. Furthermore, 
$$
\hat h_{ij}=p^{-1/5}\hat b_{ij},
$$
where $\hat b_{ij}$ is the usual CV bandwidth for the data in the
$j$th group of the $i$th partitioning. 

Because the data are independent and identically distributed, 
\beq\label{varhbar}
\Var(\bar h)&=&\frac{1}{N}\Var(\hat
h_1)+\left(\frac{N-1}{N}\right)\Cov(\hat h_1,\hat h_2)\nonumber\\
&=&\frac{1}{N}\Var(\hat
h_1)+\left(\frac{N-1}{N}\right)\Cov(\hat h_{11},\hat h_{21}).
\eeq
Let $m=n/p$ and without loss of generality let the data from which
$\hat h_{11}$ is calculated be $X_1,\ld,X_m$. The data from which
$\hat h_{21}$ is calculated are $X_{i_1},\ld,X_{i_m}$, where
$i_1,\ld,i_m$ are a random sample (without replacement) from
$1,\ld,n$. We may write
$$
\Cov(\hat h_{11},\hat h_{21})=\sum_{r=1}^m\Cov(\hat h_{11},\hat h_{21}|A_r)p_r,
$$
where $A_r$ is the event that exactly $r$ of
$i_1,\ld,i_m$ are in $\{1,\ld,m\}$ and 
$$
p_r=P(A_r)=\frac{{m\choose r}{n-m\choose
  m-r}}{{n\choose m}}.  
$$ 
Because the data are independent of the chosen permutation, 
$$
\Cov(\hat h_{11},\hat h_{21}|A_r)=p^{-2/5}\Cov(\hat b_{11},\hat b^{(r)}),
$$
where $\hat b_{11}$ and $\hat b^{(r)}$ are the usual CV bandwidths computed from
$X_1,\ld,X_m$ and $\Y=(X_1,\ld,X_r,X_{m+1},\ld,X_{2m-r})$, respectively.   

Let $CV_{11}(b)$ and $CV_r(b)$ be the cross-validation curves for
$X_1,\ld,X_m$ and $\Y$, respectively. Arguing as in the proof of
\cite{scottterrell}, as $m\ra\infty$
$$
\Cov(\hat b_{11},\hat b^{(r)})\sim
\Cov(CV'_{11}(b_0),CV'_r(b_0))\left[M''(b_0)\right]^{-2}, 
$$ 
where $b_0$ is the MISE optimal bandwidth for a sample of size $m$ and
$M''(b)$ is the usual first order approximation of the second
derivative of MISE. Arguing as in the proof of Theorem 1
$$
\Cov(CV'_{11}(b_0),CV'_r(b_0))\sim
$$
$$
4b_0^{-2}m^{-2}(m-1)^{-2}\Cov\left(\sum_{i=1}^m\sum_{j=1}^m\frac{1}{b_0}V\left(\frac{X_i-X_j}{b_0}\right),\sum_{j=1}^m\sum_{k=1}^m\frac{1}{b_0}V\left(\frac{X_{i_j}-X_{i_k}}{b_0}\right)\right),
$$
where $V$ is defined by (\ref{eq:V}). 

We may write 
$$
\sum_{i=1}^m\sum_{j=1}^m\frac{1}{b_0}V\left(\frac{X_i-X_j}{b_0}\right)=S_r+\delta_{r1}
$$
and
$$
\sum_{j=1}^m\sum_{k=1}^m\frac{1}{b_0}V\left(\frac{X_{i_j}-X_{i_k}}{b_0}\right)=S_r+\delta_{r2},  
$$
where
$$
S_r=\sum_{i=1}^r\sum_{j=1}^r\frac{1}{b_0}V\left(\frac{X_i-X_j}{b_0}\right),
$$
\beqn
\delta_{r1}&=&\sum_{i=1}^r\sum_{j=r+1}^m\frac{1}{b_0}V\left(\frac{X_i-X_j}{b_0}\right)+\sum_{i=r+1}^m\sum_{j=1}^r\frac{1}{b_0}V\left(\frac{X_i-X_j}{b_0}\right)\\
&&+\sum_{i=r+1}^m\sum_{j=r+1}^m\frac{1}{b_0}V\left(\frac{X_i-X_j}{b_0}\right)
\eeqn
and 
\beqn
\delta_{r2}&=&\sum_{i=1}^r\sum_{j=m+1}^{2m-r}\frac{1}{b_0}V\left(\frac{X_i-X_j}{b_0}\right)+\sum_{i=m+1}^{2m-r}\sum_{j=1}^r\frac{1}{b_0}V\left(\frac{X_i-X_j}{b_0}\right)\\
&&+\sum_{i=m+1}^{2m-r}\sum_{j=m+1}^{2m-r}\frac{1}{b_0}V\left(\frac{X_i-X_j}{b_0}\right).
\eeqn
It follows that 
$$
\Cov(CV'_{11}(b_0),CV'_r(b_0))\sim
$$
\beq\label{Cov}
&&4b_0^{-2}[m(m-1)]^{-2}\Bigg[\Var(S_r)+2\Cov\Bigg(S_r,\sum_{i=1}^r\sum_{j=m+1}^{2m-r}\frac{1}{b_0}V\left(\frac{X_i-X_j}{b_0}\right)\Bigg)\nonumber\\
&&+2\Cov\Bigg(S_r,\sum_{i=1}^r\sum_{j=r+1}^{m}\frac{1}{b_0}V\left(\frac{X_i-X_j}{b_0}\right)\Bigg)+\Cov(\delta_{r1},\delta_{r2})\Bigg].
\eeq

Again using \cite{scottterrell}, 
$$
\Var(S_r)=2r(r-1)b_0^{-1}\left[\int V^2(u)\,du\int f^2(x)\,dx+o(1)\right]+4r(r-1)(r-2)o(b_0^6),
$$ 
where both $o$ terms immediately above are independent of $r$. By
inspection 
$$
\Cov\left(S_r,\sum_{i=1}^r\sum_{j=m+1}^{2m-r}\frac{1}{b_0}V\left(\frac{X_i-X_j}{b_0}\right)\right)=\Cov\left(S_r,\sum_{i=1}^r\sum_{j=r+1}^{m}\frac{1}{b_0}V\left(\frac{X_i-X_j}{b_0}\right)\right).
$$
The latter covariance is 
$$
b_0^{-2}\sum_{i=1}^r\sum_{j=1}^r\sum_{k=1}^r\sum_{\ell=r+1}^m\Cov\left(V\left(\frac{X_i-X_j}{b_0}\right),V\left(\frac{X_k-X_\ell}{b_0}\right)\right)=
$$
$$
2b_0^{-2}\sum_{i=1}^r\sum_{j=1}^r\sum_{\ell=r+1}^m\Cov\left(V\left(\frac{X_i-X_j}{b_0}\right),V\left(\frac{X_i-X_\ell}{b_0}\right)\right)=
$$
$$
2b_0^{-2}r(r-1)(m-r)\Cov\left(V\left(\frac{X_1-X_2}{b_0}\right),V\left(\frac{X_1-X_3}{b_0}\right)\right).
$$

Now, 
\beqn
\Cov(\delta_{r1},\delta_{r2})&=&4b_0^{-2}\Cov\left(\sum_{i=1}^r\sum_{j=r+1}^{m}V\left(\frac{X_i-X_j}{b_0}\right),\sum_{i=1}^r\sum_{j=m+1}^{2m-r}V\left(\frac{X_i-X_j}{b_0}\right)\right)\\
&=&4b_0^{-2}r(m-r)^2\Cov\left(V\left(\frac{X_1-X_2}{b_0}\right),V\left(\frac{X_1-X_3}{b_0}\right)\right).
\eeqn
We see then that $\Cov(\delta_{r1},\delta_{r2})$ and the sum of the other
two covariance terms in (\ref{Cov}) are identical except for the
factors $r(m-r)^2$ and $r(r-1)(m-r)$. For later reference note that 
$$
\sum_{r=1}^mr(m-r)^2p_r=\frac{m^4}{n}\left(1+o(1)\right) 
$$
and 
$$
\sum_{r=1}^mr(r-1)(m-r)p_r\sim\frac{m^5}{n^2}.
$$

The sum of the first and second covariance terms in (\ref{Cov})
contributes the following to $\Cov(\hat h_{11},\hat h_{21})$: 
$$
\frac{32}{p^{2/5}m^4b_0^4M''(b_0)^2}\Cov\left(V\left(\frac{X_1-X_2}{b_0}\right),V\left(\frac{X_1-X_3}{b_0}\right)\right)\sum_{r=1}^mr(r-1)(m-r)p_r\sim
$$ 
$$
C_1m^{8/5}p^{-2/5}m^{-4}\frac{m^5}{n^2}\Cov\left(V\left(\frac{X_1-X_2}{b_0}\right),V\left(\frac{X_1-X_3}{b_0}\right)\right),
$$
where $C_1$ is a positive constant. We have
\beqn
E\left[V\left(\frac{X_1-X_2}{b_0}\right)\right]&=&\int\int
V\left(\frac{x-y}{b_0}\right)f(x)f(y)\,dx dy\\
&=&b_0\int f(x)\int V(u)f(x-b_0u)\,du dx.
\eeqn
Since $\int u^jV(u)\,du=0$, $j=0,1,2,3$, and $f$ has first four
derivatives that are bounded and continuous, 
$$
E\left[V\left(\frac{X_1-X_2}{b_0}\right)\right]=O(b_0^5).
$$
Similarly,
$$
E\left[V\left(\frac{X_1-X_2}{b_0}\right)
  V\left(\frac{X_1-X_3}{b_0}\right)\right]=O(b_0^{10}),
$$
and hence
$\Cov\left(V\left(\frac{X_1-X_2}{b_0}\right),V\left(\frac{X_1-X_3}{b_0}\right)\right)=O(b_0^{10})$.  
Therefore, the covariance term in
question contributes a term of the following order to $\Cov(\hat
h_{11},\hat h_{21})$:  
\beq\label{covcon}
m^{8/5}p^{-2/5}\frac{m}{n^2}m^{-2}=p^{-2/5}\frac{m^{3/5}}{n^{2}}.
\eeq

The term in (\ref{Cov}) involving $\Var(S_r)$ contributes the
following to $\Cov(\hat h_{11},\hat h_{21})$: 
$$
\frac{4}{p^{2/5}m^4b_0^2M''(b_0)^2}\Bigg[\frac{2}{b_0}\int V^2(u)\,du
  \int f^2(x)\,dx\sum_{r=1}^mr(r-1)p_r
$$
$$
+o(b_0^6)\sum_{r=1}^mr(r-1)(r-2)p_r\Bigg],
$$
which is 
\beq\label{varsr}
\frac{4}{p^{2/5}m^4b_0^2M''(b_0)^2}\left[\frac{2m^4}{b_0n^2}\int V^2(u)\,du
  \int f^2(x)\,dx+o(m^4b_0^{-1}n^{-2})+o(b_0^6)m^6n^{-3}\right].
\eeq
We have 
$$
o(b_0^6)m^6n^{-3}\left[\frac{m^4}{b_0n^2}\right]^{-1}=o(b_0^7)\frac{m^2}{n}=o(m/n),
$$
which tends to 0 as $m,n\ra \infty$ since $m=o(n)$. So, the
the term involving (\ref{varsr})
is asymptotic to 
\beq\label{varsr1}
\frac{8\int V^2(u)\,du\int
  f^2(x)\,dx}{p^{2/5}b_0^3M''(b_0)^2n^2}=\frac{1}{p}AV(\hat h_1),
\eeq
where $AV(\hat h_1)$ is the asymptotic variance of $\hat h_1$. The
order of (\ref{varsr1}) is $m^{7/5}p^{-2/5}n^{-2}$, and hence larger
than (\ref{covcon}).

The term in (\ref{Cov}) involving $\Cov(\delta_{r1},\delta_{r2})$
contributes the following to $\Cov(\hat h_{11},\hat h_{21})$: 
$$
\frac{16}{p^{2/5}m^4b_0^4M''(b_0)^2}\Cov\left(V\left(\frac{X_1-X_2}{b_0}\right),V\left(\frac{X_1-X_3}{b_0}\right)\right)\sum_{r=1}^mr(m-r)^2p_r,
$$ 
which is of order 
$m^{8/5}m^{-4}p^{-2/5}m^{-2}m^4n^{-1}=(pm)^{-2/5}n^{-1}=n^{-7/5}$.

Collecting previous results, we have
\beq\label{covh11h21}
\Cov(\hat h_{11},\hat h_{21})=\frac{1}{p}AV(\hat
h_1)+O(n^{-7/5})+o\left(\frac{1}{p}AV(\hat h_1)\right). 
\eeq
Since $p^{-1}AV(\hat h_1)$ is asymptotic to $C_4p^{-9/5}n^{-3/5}$, it
follows that $n^{-7/5}$ is of smaller order than $p^{-1}AV(\hat
h_1)$ when $p=o(n^{4/9})$.  Combining (\ref{covh11h21}) with
(\ref{varhbar}), and assuming that $N\sim p^a$ for $a>1$, we have
$$
\Var(\bar h)\sim \frac{1}{p}AV(\hat h_1).
$$  
On the other hand, if $N$ is fixed, then $\Var(\bar h)$ is
asymptotic to $\Var(\hat h_1)/N$ as $n$ and $p$ tend to $\infty$.

\section{Appendix B}

{\bf Proof of Lemma \ref{lem:U_stat}.} We prove the first equality; the rest follow similarly. We have
\begin{align*}
\int \hat{f}_h^2(x) dx 
& = \frac{1}{n^2 h^2} \sum_{i=1}^n \sum_{j=1}^n \int K\bigg( \frac{x - X_i}{h} \bigg) K\bigg( \frac{x - X_j}{h} \bigg) dx\\
& = \frac{1}{n^2 h} \sum_{i=1}^n \sum_{j=1}^n \int K\bigg( \frac{X_i - X_j}{h} - u \bigg) K(u) du \\
& = \frac{1}{n^2h} \sum_{i=1}^n \sum_{j=1}^n A \bigg( \frac{X_i - X_j}{h} \bigg).
\end{align*}

\biblist

\end{document}